\UseRawInputEncoding
\documentclass[superscriptaddress, twocolumn, amsmath, amssymb, aps,prl, notitlepage,longbibliography,10pt]{revtex4-1} 
\usepackage{graphicx,graphics,epsfig,subfigure,times,bm,bbm,amssymb,amsmath,amsfonts,amsthm,mathrsfs,MnSymbol,physics}
\usepackage[matrix,frame,arrow]{xypic}
\usepackage[normalem]{ulem}
\usepackage{slashed}
\usepackage{dcolumn}
\usepackage{tabularx}
\usepackage{color}
\usepackage[usenames,dvipsnames,svgnames,table]{xcolor}
\usepackage[english]{babel}
\usepackage{verbatim}
\definecolor{darkblue}{rgb}{0.0,0.0,0.3}
\usepackage[colorlinks=true,
            linkcolor=red,
            urlcolor= darkblue,
            citecolor=blue]{hyperref}

\newcommand{\bea}{\begin{eqnarray}}
\newcommand{\eea}{\end{eqnarray}}

\begin{document}
\title{Global-Local Duality of Energetic Control Cost in Multipartite Quantum Correlated Systems}

\author{Rui Guan}
\address{QianWeiChang College, Shanghai University, Shanghai, 200444, China}
\author{Junjie Liu}
\email{jj\_liu@shu.edu.cn}
\affiliation{Institute for Quantum Science and Technology, Shanghai Key Laboratory of High Temperature Superconductors, Department of Physics, International Center of Quantum and Molecular Structures, Shanghai University, Shanghai, 200444, China}
\address{QianWeiChang College, Shanghai University, Shanghai, 200444, China}
\author{Jian-Hua Jiang}
\email{jhjiang3@ustc.edu.cn}
\affiliation{School of Biomedical Engineering, Division of Life Sciences and Medicine, University of Science and Technology of China, Hefei 230026, China}
\affiliation{Suzhou Institute for Advanced Research, University of Science and Technology of China, Suzhou, 215123, China}
\affiliation{School of Physical Sciences, University of Science and Technology of China, Hefei, 230026, China}

\begin{abstract}
Multipartite quantum correlated systems (MQCSs) are widely utilized in diverse quantum information tasks, where their sophisticated control inherently incurs energetic costs. However, the fundamental characteristics of these control costs remain elusive, largely due to the lack of thermodynamic descriptions capable of capturing the full complexities of MQCSs. Here, we uncover universal thermodynamic relations for arbitrary MQCSs weakly coupled to a thermal bath, establishing an intrinsic global-local duality of control costs. Using these relations, we elucidate the exact role of multipartite correlation--a defining quantum feature of MQCSs--in shaping control costs at finite times. We also demonstrate that the relative magnitude between global and local control costs is undetermined, which perplexes the cost management of MQCSs under finite-time controls. Our results are numerically corroborated with applications to experimentally realizable multi-qubit systems undergoing finite-time qubit reset processes. 
\end{abstract}

\date{\today}

\maketitle

{\it Introduction}---Realizing quantum technologies that outperform their classical counterparts requires careful energetic considerations~\cite{Banacloche.02.PRL,Auffeves.22.PRXQ}. Since quantum tasks are generally carried out by dynamical quantum systems, they proceed along with changes in the energetic and entropic characteristics of these systems. Portraying this intertwined feature of quantum tasks necessitates an interdisciplinary perspective, exemplified by the pursuit of connections between information and energy for understanding the intersection between information science and thermodynamics, as pioneered by Maxwell, Szilard and Landauer and proceeded with many others~\cite{Szilard.29.ZP,Landauer.61.IBM,Maruyama.09.RMP,Parrondo.15.NP,Goold.16.JPA,LiuJ.24.PRR}. Indeed, the advancement of quantum technologies can benefit from this interdisciplinary perspective, which not only reveals the constraints on the performance of quantum tasks imposed by fundamental thermodynamic laws~\cite{Landauer.61.IBM,Maruyama.09.RMP,Landi.21.RMP,Danageozian.22.PRXQ,Parrondo.15.NP,Goold.16.JPA,LiuJ.24.PRR}, but also facilitates the study of the emerging issue of their associated thermodynamic cost~\cite{Banacloche.02.PRL,Auffeves.22.PRXQ,Nielsen.98.PRSL,Sagawa.09.PRL,Horowitz.15.PRL,Boyd.18.PRX,Abah.19.NJP,Deffner.21.EPL,Pearson.21.PRX,Chiribella.22.NC,Woods.23.PRX}.    

Multipartite quantum correlated systems (MQCSs), which hold significant promise for quantum information applications ~\cite{Bloch.08.RMP,Chin.12.PRL,Georgescu.14.RMP,Adesso.16.JPA,Koch.16.JPA,Chiara.18.RPP,Shahandeh.19.PRA,Narang.20.CR,Britton.12.N,Labuhn.16.N,Monroe.21.RMP,Periwal.21.N,Mivehvar.21.AP,Zwolak.23.RMP,MiX.24.S,Defenu.23.RMP,Feng.23.N}, provide compelling platforms to integrate the principles of quantum information science and quantum thermodynamics. In this regard, different exotic aspects pertaining to quantum correlation in quantum thermodynamics have been reported in the literature which greatly expand the structure of quantum thermodynamics, notable examples include refined thermodynamic relations~\cite{Sagawa.12.PRL,Bera.17.NC}, correlation-involved resource theory~\cite{Sapienza.19.NC}, correlation-assisted work extraction~\cite{Oppenheim.02.PRL,Llobet.15.PRX,Manzano.18.PRL} and correlation-induced anomalous energy transfer~\cite{Micadei.19.NC,Bartosik.24.PRL} to name just a few. In contrast, the energetic cost of controlling MQCSs--termed control cost from here on--remains poorly understood, despite the recognition that MQCSs, when operated with well-designed control protocols, can advance the potential of various quantum information tasks \cite{Bloch.08.RMP,Britton.12.N,Labuhn.16.N,Monroe.21.RMP,Periwal.21.N,Mivehvar.21.AP,Zwolak.23.RMP,MiX.24.S,Defenu.23.RMP,Feng.23.N}. Existing studies in this direction have largely focused on either closed systems~\cite{Huber.15.NJP} or simplified open bipartite settings under additional assumptions~\cite{Rio.11.N,Boyd.18.PRX}, leaving a general understanding of thermodynamic control cost in generic MQCSs still lacking.

\begin{figure}[b!]
 \centering
\includegraphics[width=0.8\columnwidth]{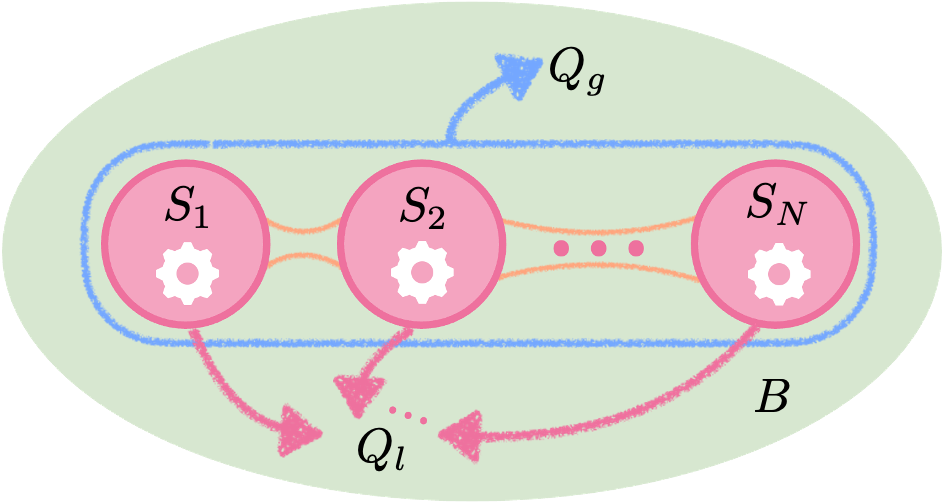} 
\caption{Schematic picture of the study. We consider MQCSs with several local parties $\{S_i\}$ that are weakly coupled to a thermal bath ($B$) and subjected to arbitrary control protocols. By examining both the global cost $Q_g$ and the local one $Q_l$, evaluated at the global and local levels respectively, we unveil general relations [cf. Eqs. (\ref{eq:sum_rule}) and (\ref{eq:low_bound})] that reflect a general global-local duality in the control cost.
  }
\protect\label{fig:scheme}
\end{figure}
We here focus on thermodynamic control cost of generic MQCSs that are immersed in a thermal bath and subjected to arbitrary finite-time control protocols (see Fig. \ref{fig:scheme} for a sketch). By combining global and local thermodynamic descriptions, we uncover fundamental relations that embody a general global-local duality of the control cost in MQCSs, requiring the minimum assumption of weak system-bath couplings. Specifically, we reveal an exact sum rule [Eq. (\ref{eq:sum_rule})] for the globally evaluated control cost, which encompasses contributions from quantum multipartite mutual information~\cite{Groisman.05.PRA,Modi.12.RMP,Huber.15.NJP,Chiara.18.RPP}--quantifying multipartite total correlation of MQCSs--and dissipated work~\cite{Crooks.99.PRE,Vaikuntanathan.09.EL}, that signifies finite-time thermodynamic processes. We thus add a previously unidentified structure to the collection of energy-information links~\cite{Szilard.29.ZP,Landauer.61.IBM,Maruyama.09.RMP,Parrondo.15.NP,Goold.16.JPA,LiuJ.24.PRR}. We also identify a general lower bound [Eq. (\ref{eq:low_bound})] on the contrast between globally and locally evaluated costs (Fig. \ref{fig:scheme}) in terms of global dissipated work and changes in intra-system interaction energy. 

Using the established relations, we elucidate the exact role of multipartite correlation in shaping control costs. Notably, the interplay between multipartite correlation and dissipated work leads to an indefinite relative magnitude between globally and locally evaluated control costs, introducing significant complexity in the thermodynamic cost management of MQCSs under finite-time control protocols. We substantiate the theoretical findings using experimentally feasible multi-qubit systems undergoing a finite-time qubit reset process. The thermodynamic framework presented here allows the description of MQCSs with arbitrary complexity, an aspect crucial for describing current experimental platforms ~\cite{Bloch.08.RMP,Britton.12.N,Labuhn.16.N,Monroe.21.RMP,Periwal.21.N,Mivehvar.21.AP,Zwolak.23.RMP,MiX.24.S,Defenu.23.RMP,Feng.23.N} and advancing the thermodynamics of control in complex quantum systems.

{\it General setup and main results.}---We consider a generic MQCS that is weakly coupled to a thermal bath at a temperature $T$ and experiences time-dependent control protocols. The Hamiltonian of the MQCS reads (Setting $\hbar=1$ and $k_B=1$ hereafter)
\begin{equation}\label{eq:hami}
    H_{g}(t)=\sum_iH_i(t)+H_{\rm{I}}(t).
\end{equation}
Here, $\{H_{i}(t)\}$ denotes the set of Hamiltonian for the driven local parties of the global system. $H_{\rm{I}}(t)$ encompasses all intra-system interaction terms, with coupling strengths that may assume any magnitudes depending on the concrete setups. Unlike previous theoretical studies that often assumed noninteracting local parties for simplicity, we consider $H_{\rm{I}}(t)\neq 0$ to align with experimental conditions. In scenarios where intra-system couplings cannot be controlled, $H_{\rm{I}}$ becomes time-independent.

The setup in Eq. (\ref{eq:hami}) not only represents a broad class of realistic many-body physical systems, but also captures processes such as quantum measurement and feedback control as special cases where effective bipartite descriptions are applicable~\cite{Sagawa.13.NJP,Barato.13.PRE,Horowitz.14.PRX}. We do not include the Hamiltonian of the bath and system-bath interaction in Eq. (\ref{eq:hami}), as they are irrelevant for getting the following main results. The evolution of the global system is described by a time-evolved density matrix $\rho_{g}(t)$, with $\rho_i(t) = \mathrm{Tr}_{\bar{i}}[\rho_{g}(t)]$ being the marginal for the $i$-th local party after tracing out degrees of freedom $\bar{i}$ complementing that of the subsystem $i$.

We decompose the change in internal energy of the global system $E_{g}(t)\equiv \mathrm{Tr}[H_{g}(t)\rho_{g}(t)]$ according to the first law of thermodynamics, $\Delta E_{g}(t)=W_{g}(t)-Q_g(t)$. Here, $W_{g}(t)\equiv\int_0^t\mathrm{Tr}[\rho_{g}(\tau)\frac{d}{d\tau}H_{g}(\tau)]d\tau$ denotes the amount of work performed on the global system by external driving fields during the time interval $[0,t]$~\cite{Bruch.16.PRB,Cangemi.21.PRR,Liu.21.PRL}, $Q_g(t)$ represents the heat dissipation flowing out of the global system which we refer to as the global control cost. For local parties, we can also decompose the energy change $\Delta E_i(t)=W_i(t)-Q_i(t)$ with $E_i(t)=\mathrm{Tr}[H_i(t)\rho_i(t)]$ and $W_{i}(t)\equiv\int_0^t\mathrm{Tr}[\rho_{i}(\tau)\frac{d}{d\tau}H_{i}(\tau)]d\tau$ once local information $\{H_i(t),\rho_i(t)\}$ is available. We refer to $Q_l(t)=\sum_iQ_i(t)$ as the local control cost as it permits a local access. We emphasize that the global and local control costs are distinguished by their reliance on global and local measurements, respectively, with the control fields in $H_g(t)$ remaining the same during evaluations.

We find that the global and local control costs are connected through an exact sum rule
\begin{equation}\label{eq:sum_rule}
    Q_g(t) = Q_l(t) +T\Delta I_g(t) + W_{\rm{dis}}^{\Delta}(t).
\end{equation}
The proof is postponed to the end of the Letter. Here, $\Delta I_g(t)=I_g(t)-I_g(0)$ denotes the change in quantum multipartite mutual information $I_g(t) \equiv \sum_i S_i(t)-S_{g}(t)$~\cite{Modi.12.RMP,Huber.15.NJP,Chiara.18.RPP} during the process. $I_g(t)$ quantifies the multipartite total correlation within the MQCS with $S_{i,g}(t)=-\mathrm{Tr}[\rho_{i,g}(t)\ln\rho_{i,g}(t)]$ the von Neumann entropy and reduces to the usual quantum mutual information in bipartite settings. We have defined $W_{\rm{dis}}^{\Delta}(t)\equiv W_{\rm{dis}}^{g}(t)-W_{\rm{dis}}^{l}(t)$ to denote the contrast between global and local dissipated work~\cite{Crooks.99.PRE,Vaikuntanathan.09.EL} which are respectively defined as $W_{\rm{dis}}^{g}(t)\equiv W_{g}(t)-\Delta F_{g}(t)$ and $W_{\rm{dis}}^{l}(t)\equiv\sum_{i}[W_i(t)-\Delta F_i(t)]$ with $\Delta F_j(t)$ ($j=i,g$) denoting the change in nonequilibrium free energy $F_j(t)=E_j(t)-TS_j(t)$~\cite{Parrondo.15.NP,Deffner.13.PRX,Liu.23.PRAa}. 
\begin{figure*}[t!]
 \centering
\includegraphics[width=2\columnwidth]{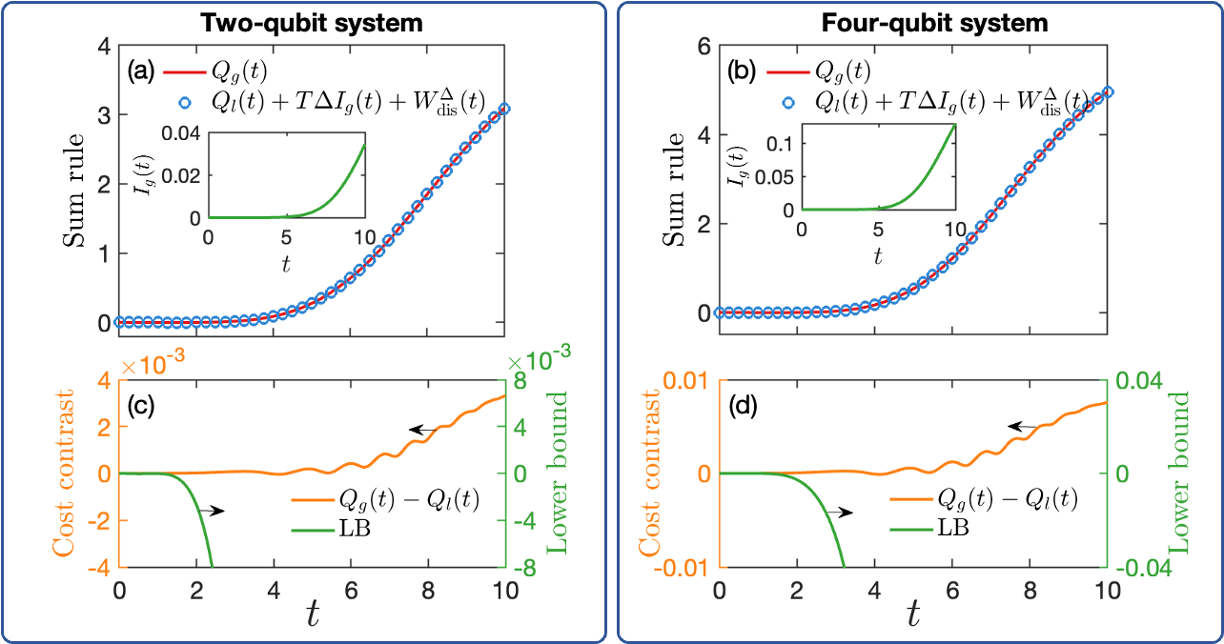} 
\caption{Examining Eqs. (\ref{eq:sum_rule}) and (\ref{eq:low_bound}) in multi-qubit systems with $N=2$ (left panel) and $N=4$ (right panel). (a)(b): Validity of the sum rule for the global cost $Q_g(t)$. (c)(d): The cost contrast $Q_g(t)-Q_l(t)$ (solid line, left axis) and its lower bound (LB) (dashed line, right axis). Insets: Time-dependent quantum multipartite mutual information $I_g(t)$. Other parameters are $\beta=1$, $\lambda=0.02$, $\gamma=0.02$, $\varepsilon_0=0.4$, $\varepsilon_{\tau}=10$ and $\tau=10$.}
\protect\label{fig:data}
\end{figure*}
Notably, Eq. (\ref{eq:sum_rule}) represents an energy-information link, strengthening the synergy between quantum thermodynamics and quantum information science in MQCSs.

The sum rule Eq. (\ref{eq:sum_rule}) indicates that the global and local costs can differ in MQCSs. Notably, we find that the second law of thermodynamics, when applied to the global system, imposes a general lower bound on the magnitude of the cost contrast $Q_g(t)-Q_l(t)$,
\begin{equation}\label{eq:low_bound}
    Q_g(t)-Q_l(t) \geqslant -W_{\rm{dis}}^g(t)-\Delta E_{\rm{I}}(t).
\end{equation}
Here, $\Delta E_{\rm I}(t)$ marks the change in the interaction energy $E_{\rm I}(t)=\mathrm{Tr}[H_{\rm I}(t)\rho_{g}(t)]$. For clarity, we provide the proof details at the end of the Letter. The equality of Eq. (\ref{eq:low_bound}) is attained in the thermodynamic reversible limit. Hence, we expect that the lower bound becomes loose for finite-time processes and cannot be utilized to infer the exact magnitude of the cost contrast.

Eqs. (\ref{eq:sum_rule}) and (\ref{eq:low_bound}) constitute our main results. We emphasize that Eqs. (\ref{eq:sum_rule}) and (\ref{eq:low_bound}) are derived without imposing assumptions on the details of the Hamiltonian in Eq. (\ref{eq:hami}). Hence they establish a general global-local duality of control costs in MQCSs, which holds across the finite-time regime to the thermodynamic reversible limit. In the special limit of $H_{\rm I}(t)=0$, we can analytically demonstrate that $Q_g(t)=Q_l(t)$, regardless of the magnitude of multipartite correlation; We relegate proof details to the Supplemental Material~\cite{SM}. In this case, the lower bound in Eq. (\ref{eq:low_bound}) reduces to the maximum work principle for the global system, $W_{\rm{dis}}^g(t)\ge 0$. However, we emphasize that $H_{\rm I}(t)\neq0$ represents the typical scenario in experiments~\cite{Britton.12.N,Labuhn.16.N,Monroe.21.RMP,Mivehvar.21.AP,Defenu.23.RMP,Feng.23.N,Periwal.21.N,Zwolak.23.RMP,MiX.24.S}, rendering Eqs. (\ref{eq:sum_rule}) and (\ref{eq:low_bound}) generally nontrivial universal relations without existing counterparts.

It is worth delineating critical properties and implications of Eqs. (\ref{eq:sum_rule}) and (\ref{eq:low_bound}). (i) Eq. (\ref{eq:sum_rule}) indicates that the interplay between multipartite correlation and dissipated work renders the cost contrast $Q_g(t)-Q_l(t)$, while bounded from below as stated in Eq. (\ref{eq:low_bound}), lacking a definite sign. This perplexes the cost management of MQCSs under finite-time operations.
(ii) By examining the asymptotic behavior of Eq. (\ref{eq:sum_rule}) in the thermodynamic reversible limit where $W_{\rm{dis}}^{g}(t)$ vanishes, we point out that the dissipated work contrast $W_{\rm{dis}}^{\Delta}(t)$ remains nonzero. This occurs because the local parties of a correlated system no longer adhere to the standard form of the second law of thermodynamics~\cite{Bera.17.NC}, as demonstrated in~\cite{SM}. (iii) We remark that the global cost $Q_g(t)$ should comply with the standard form of the Landauer principle, $Q_g(t)\ge -T\Delta S_g(t)$~\cite{Landi.21.RMP}. In comparison, by combining this inequality with Eq. (\ref{eq:sum_rule}), we find that the local cost satisfies a modified form of the Landauer principle, $Q_l(t)\ge -T\sum_i\Delta S_i(t)-W_{\rm{dis}}^{\Delta}(t)\neq -T\sum_i\Delta S_i(t)$ in light of the second point mentioned above. (iv) Finally, we note that maintaining multipartite correlation in MQCSs over time is essential for unlocking their scalability advantages. In this regard, Eqs. (\ref{eq:sum_rule}) and (\ref{eq:low_bound}) translate into thermodynamic constraints on the magnitude of $\Delta I_g(t)$ that can be actively achieved in MQCSs. Specifically, Eq. (\ref{eq:sum_rule}) implies that tuning $\Delta I_g(t)$ necessarily induces global-local contrasts in control cost and dissipated work. Combining Eqs. (\ref{eq:sum_rule}) and (\ref{eq:low_bound}), we find that $\Delta I_g(t)$ is subject to a thermodynamic lower bound, $T\Delta I_g(t)\ge W_{\rm{dis}}^l(t)-2W_{\rm{dis}}^g(t)-\Delta E_{\rm I}(t)$.

{\it Numerical demonstration.}---We validate our main theoretical findings using multi-qubit systems weakly coupled to a thermal bath at a temperature $T=\beta^{-1}$ with open boundaries
\begin{equation}\label{eq:HS}
    H_g(t) = \frac{\varepsilon_t}{2}\sum_{i=1}^N\Big[\cos (\theta_t)\sigma_i^{z}+\sin(\theta_t)\sigma_i^{x}\Big]+\sum_{i=1}^{N-1}\lambda_t \sigma_i^{x}\sigma_{i+1}^{x}.
\end{equation}
Here, $N$ marks the number of qubits, $\{\sigma_i^{x,z}\}$ represent the Pauli matrices, $\varepsilon_t$, $\theta_t$ and $\lambda_t$ are subject to time-dependent control protocols. The multi-qubit model with nearest-neighbor couplings generalizes a single-qubit setup~\cite{Miller.20.PRL,Riechers.21.PRA,Liu.23.PRAa}, and can be experimentally realized using coupled quantum dot systems~\cite{Kandel.19.N} and tight-binding superconducting qubit arrays~\cite{Karamlou.22.NPJQ}, where both the local parties and couplings are tunable. We here consider the following finite-time control fields $\varepsilon_t = \varepsilon_0 +(\varepsilon_\tau - \varepsilon_0)\sin (\pi t/2\tau)^2$, $\theta_t=\frac{\pi}{2}(t/\tau-1)$ with $\tau$ setting the operation time limit and $\lambda_t=\lambda\cos(\theta_t)$. We take these experimentally feasible control fields~\cite{Miller.20.PRL,Riechers.21.PRA,Liu.23.PRAa} to implement a multi-qubit reset process, a procedure with significant applications in numerous quantum algorithms~\cite{DiVincenzo.00.FP} and quantum error correction~\cite{Schindler.11.S,Reed.12.N}. 

At weak couplings, the evolution of $\rho_{g}(t)$ is governed by the quantum Lindblad master equation   
\begin{equation}\label{eq:master_equation}
    \dv{t} \rho_{g}(t)=-i[H_g(t),\rho_{g}(t)]+\sum_{\mu=1,2} \gamma_{\mu}\mathcal{D}[L_{\mu}]\rho_{g}(t).
\end{equation}
Here, $\gamma_{\mu} \geqslant 0$ is the damping coefficient of quantum channel $\mu$, and $\mathcal{D}[L_{\mu}]\rho = L_{\mu}\rho L_{\mu}^{\dagger}-\frac{1}{2}\left\{L_{\mu}^{\dagger}L_{\mu},\rho\right\}$ is the Lindblad superoperator with $L_{\mu}$ being the Lindblad jump operator and $\left\{A, B\right\}= AB + BA$. For the present model, the Lindblad jump operators due to the coupling to a common thermal bath read~\cite{Cattaneo.19.NJP}
$L_1(t)=\sqrt{\varepsilon(t)[N_B(t)+1]}\sum_i\sigma_i^{-}$ and $L_2(t)=\sqrt{\varepsilon(t)[N_B(t)]}\sum_i\sigma_i^{+}$, where $\sigma_i^{+,-}$ denote spin ladder operators of the $i$th qubit and $N_B(t)=[e^{\beta\varepsilon(t)}-1]^{-1}$. For the sake of simplicity, we take $\gamma_{1,2}=\gamma$ in simulations. A time step $\delta t=1\times 10^{-4}$ is used to numerically integrate Eq. (\ref{eq:master_equation}).

For the demonstration purpose, we choose an initial thermal state $\rho_{g}(0)= e^{-\beta H_{g}(0)}/\mathrm{Tr}[e^{\beta H_{g}(0)}]$ for the global system and depict a typical set of numerical results in Fig. \ref{fig:data}. In Fig. \ref{fig:data} (a) and (b) for $N=2$ and $N=4$, respectively, we show numerical results for quantities on two sides of Eq. (\ref{eq:sum_rule}). The perfect agreement between both clearly demonstrates the validity of the sum rule in multi-qubit systems. From the result, we note that the global cost $Q_g(t)$ monotonically increases as the system evolves. In Fig. \ref{fig:data} (c) and (d), we verify Eq. (\ref{eq:low_bound}) by presenting numerical results for the cost contrast $Q_g(t)-Q_l(t)$ and its lower bound (LB) as given by the right-hand-side of Eq. (\ref{eq:low_bound}). We observe that $Q_g(t)-Q_l(t)>0$ especially at large times from both Fig. \ref{fig:data} (c) and (d). 

To analyze the sign of $Q_g(t)-Q_l(t)$ in the present model, we rewrite the sum rule as $Q_g(t)-Q_l(t)=T\Delta I_g(t)+W^{\Delta}(t)-F^{\Delta}(t)$, where we denote $F^{\Delta}(t)\equiv \Delta F_g(t)-\sum_{i}\Delta F_i(t)$ and $W^{\Delta}(t)\equiv W_g(t)-\sum_{i}W_i(t)$. Specifically, $F^{\Delta}(t)$ represents the contrast in work capacity (i.e., free energy change) between the global system and its local parties. Since intra-system correlations act as a thermodynamic resource that enables additional work extraction~\cite{Rio.11.N,Llobet.15.PRX}, it is natural to expect that the global system has a greater work capacity than its local parties, leading to $F^{\Delta}(t)>0$ as confirmed numerically~\cite{SM}. The term $W^{\Delta}(t)$ becomes nonzero when the intra-system interaction is actively controlled. In our model, we continuously increase the strength of the intra-system interaction, requiring an additional work performed on the system such that $W^{\Delta}(t)>0$~\cite{SM}. Additionally, we observe $\Delta I_g(t)>0$ from the insets of Fig. \ref{fig:data} (a) and (b), indicating that control fields generate multipartite correlations. Since creating correlations inherently incurs an energetic cost as exactly quantified by $T\Delta I_g(t)$~\cite{Huber.15.NJP}, we can attribute the sign of $Q_g(t)-Q_l(t)$ here to the interplay between the gain in work capacity $F^{\Delta}(t)$ and energetic costs of building up correlation and interaction as captured by $T\Delta I_g(t)+W^{\Delta}(t)$. 

For the scenario considered in Fig. \ref{fig:data}, we find that the costs outweigh the gain, leading to a positive cost contrast $Q_g(t)-Q_l(t)>0$. However, when the gain instead surpasses the costs, the scenario of $Q_g(t)-Q_l(t)<0$ can also occur, as demonstrated using a multi-qubit model with a modified intra-system interaction and the same control fields (see details in~\cite{SM}). The varying signs of $Q_g(t)-Q_l(t)$ in multi-qubit systems confirm our expectations that the cost contrast lacks a definite sign, and that the sum rule captures the rich dynamical behaviors of control costs in MQCSs.

{\it Proof of Eqs. (\ref{eq:sum_rule}) and (\ref{eq:low_bound}).}---To proof Eq. (\ref{eq:sum_rule}), we start from the nonequilibrium free energy~\cite{Parrondo.15.NP,Deffner.13.PRX,Liu.23.PRAa} of the global system, $F_{g}(t)~=~E_{g}(t)-TS_{g}(t)$. Combining the first law of thermodynamics for the global system and the definition of quantum multipartite mutual information~\cite{Modi.12.RMP,Huber.15.NJP,Chiara.18.RPP}, the change in $F_g(t)$ can be expressed as $\Delta F_{g}(t)=W_{g}(t)-Q_g(t)-T[\sum_i\Delta S_i(t)-\Delta I_g(t)]$. To proceed, we further consider nonequilibrium free energies $F_i(t)~=~E_i(t)-TS_i(t)$ for local parties which can be adopted to replace the term $\sum_i\Delta S_i(t)$ in $\Delta F_g(t)$, yielding 
the following relation
\bea\label{eq:ddf}
    Q_g(t) &=& W_{g}(t)-\Delta F_{g}(t)+T\Delta I_g(t)\nonumber\\
    &&+\sum_i[\Delta F_i(t)-\Delta E_i(t)].
\eea
The sum rule Eq. (\ref{eq:sum_rule}) can be immediately derived from Eq. (\ref{eq:ddf}) by invoking the energy decomposition for local parties $\Delta E_i(t)=W_i(t)-Q_i(t)$ and applying the definition of dissipated work $W_{\rm dis}\equiv W-\Delta F$~\cite{Crooks.99.PRE,Vaikuntanathan.09.EL} to both the global system and local parties.

We then turn to the cost contrast $Q_g(t)-Q_l(t)$. Inserting the energy decomposition $E_{g}(t)=\sum_{i} E_{i}(t)+ E_{\rm{I}}(t)$ into the nonequilibrium free energy $F_{g}(t)=E_{g}(t)-TS_{g}(t)$ of the global system and considering the latter's change, we get 
\begin{equation}\label{eq:s1}
    T\Delta S_{g}(t)~=~\sum_{i}\Delta E_{i}(t)+\Delta E_{\rm{I}}(t)-\Delta F_{g}(t).
\end{equation}
We next invoke the relation $\Delta E_i(t)=W_i(t)-Q_i(t)$ and transform Eq. (\ref{eq:s1}) into
\begin{equation}\label{eq:c2}
        T\Delta S_{g}(t)~=~-Q_l(t)+\Delta E_{\rm{I}}(t)+ W_{\rm{dis}}^g(t).
\end{equation}
To arrive at the above form, we have utilized the fact that $W_{g}(t)= \sum_i W_i(t)$ since only the local system Hamiltonian $\{H_i(t)\}$ is being driven. 

To proceed, we further add $Q_g(t)$ to both sides of Eq. (\ref{eq:c2}) and utilize the second law of thermodynamics satisfied by the global system $T\Delta S_{g}(t) +Q_g(t)\geqslant 0$, yielding
\begin{equation}
    Q_g(t)-Q_l(t) +\Delta E_{\rm{I}}(t)+ W_{\rm{dis}}^g(t)~\geqslant~0.
\end{equation}
By rearranging the terms, we arrive at Eq. (\ref{eq:low_bound}). From the above derivations, we clearly see that Eqs. (\ref{eq:sum_rule}) and (\ref{eq:low_bound}) are general.

{\it Discussion and conclusion.}---The experimental verifications of Eqs. (\ref{eq:sum_rule}) and (\ref{eq:low_bound}), though challenging due to the necessity of global measurements on $\rho_g(t)$, is well within the reach of current experimental capabilities. We remark that the detailed form of $H_g(t)$ is known {\it a priori} in typical experiments~\cite{Britton.12.N,Labuhn.16.N,Monroe.21.RMP,Periwal.21.N,Mivehvar.21.AP,Zwolak.23.RMP,MiX.24.S,Defenu.23.RMP,Feng.23.N}. $\rho_g(t)$ can be measured using quantum state tomography techniques advanced for many-body systems such as qubit arrays~\cite{Lanyon.17.NP,Yoneda.21.NC,Zhong.21.N}, whose capability can be further enhanced by incorporating machine learning methods~\cite{Torlai.18.NP, Ahmed.21.PRL,Struchalin.21.PRXQ}.

We note that our main results [cf. Eqs. (\ref{eq:sum_rule}) and (\ref{eq:low_bound})] require the existence of a fixed bath temperature. However, in realistic nanoscale environments, which are often finite in size, time-dependent effective temperatures provide a more accurate description~\cite{Schaller.14.NJP,Amato.20.PRA,Campeny.21.PRXQ,Pekola.21.RMP,Yuan.22.PRE,Spiecker.23.NP,Moreira.23.PRL}. To accommodate such situations, we can adopt a generalized definition of the nonequilibrium free energy~\cite{Liu.23.PRAa} which can incorporate a time-dependent effective temperature. We show that the basic form of the sum rule remains intact, with just the term $T\Delta I_g(t)$ in Eq. (\ref{eq:sum_rule}) replaced by $T(t)I_g(t)-T(0)I_g(0)$. However, the lower bound on $Q_g(t)-Q_l(t)$ as given by Eq. (\ref{eq:low_bound}) cannot be straightforwardly generalized (see details in~\cite{SM}). Thus we expect that at least the global-local duality underpinning the sum rule may represent a general feature of a wide range of nonequilibrium MQCSs, without coupling to a thermal bath.

In summary, we provided a general thermodynamic description for addressing the control cost of arbitrary MQCSs under finite-time operations. We uncover a fundamental global-local duality governing the control cost of MQCSs, elucidating the interplay between relevant thermodynamic and information-theoretic quantities. The critical properties and implications of our results are delineated and demonstrated using experimentally feasible multi-qubit models. We believe that these findings offer fresh insights into the thermodynamics of control in MQCSs, which are particularly relevant for applications of MQCSs in quantum simulation, error correction and beyond.

{\it Data availability statement}--The data to generate Figure 2 are available upon reasonable request from the authors.

{\it Acknowledgments}---J.L. acknowledges support from the National Natural Science Foundation of China (Grant No. 12205179), the Shanghai Pujiang Program (Grant No. 22PJ1403900) and start-up funding of Shanghai University. J.-H. Jiang acknowledges support from the National Natural Science Foundation of China (Grant No. 12125504) and the ``Hundred Talents Program'' of the Chinese Academy of Sciences.


\begin{thebibliography}{80}%
\makeatletter
\providecommand \@ifxundefined [1]{%
 \@ifx{#1\undefined}
}%
\providecommand \@ifnum [1]{%
 \ifnum #1\expandafter \@firstoftwo
 \else \expandafter \@secondoftwo
 \fi
}%
\providecommand \@ifx [1]{%
 \ifx #1\expandafter \@firstoftwo
 \else \expandafter \@secondoftwo
 \fi
}%
\providecommand \natexlab [1]{#1}%
\providecommand \enquote  [1]{``#1''}%
\providecommand \bibnamefont  [1]{#1}%
\providecommand \bibfnamefont [1]{#1}%
\providecommand \citenamefont [1]{#1}%
\providecommand \href@noop [0]{\@secondoftwo}%
\providecommand \href [0]{\begingroup \@sanitize@url \@href}%
\providecommand \@href[1]{\@@startlink{#1}\@@href}%
\providecommand \@@href[1]{\endgroup#1\@@endlink}%
\providecommand \@sanitize@url [0]{\catcode `\\12\catcode `\$12\catcode
  `\&12\catcode `\#12\catcode `\^12\catcode `\_12\catcode `\%12\relax}%
\providecommand \@@startlink[1]{}%
\providecommand \@@endlink[0]{}%
\providecommand \url  [0]{\begingroup\@sanitize@url \@url }%
\providecommand \@url [1]{\endgroup\@href {#1}{\urlprefix }}%
\providecommand \urlprefix  [0]{URL }%
\providecommand \Eprint [0]{\href }%
\providecommand \doibase [0]{http://dx.doi.org/}%
\providecommand \selectlanguage [0]{\@gobble}%
\providecommand \bibinfo  [0]{\@secondoftwo}%
\providecommand \bibfield  [0]{\@secondoftwo}%
\providecommand \translation [1]{[#1]}%
\providecommand \BibitemOpen [0]{}%
\providecommand \bibitemStop [0]{}%
\providecommand \bibitemNoStop [0]{.\EOS\space}%
\providecommand \EOS [0]{\spacefactor3000\relax}%
\providecommand \BibitemShut  [1]{\csname bibitem#1\endcsname}%
\let\auto@bib@innerbib\@empty
\bibitem [{\citenamefont {Gea-Banacloche}(2002)}]{Banacloche.02.PRL}%
  \BibitemOpen
  \bibfield  {author} {\bibinfo {author} {\bibfnamefont {J.}~\bibnamefont
  {Gea-Banacloche}},\ }\bibfield  {title} {\enquote {\bibinfo {title} {Minimum
  energy requirements for quantum computation},}\ }\href {\doibase
  10.1103/PhysRevLett.89.217901} {\bibfield  {journal} {\bibinfo  {journal}
  {Phys. Rev. Lett.}\ }\textbf {\bibinfo {volume} {89}},\ \bibinfo {pages}
  {217901} (\bibinfo {year} {2002})}\BibitemShut {NoStop}%
\bibitem [{\citenamefont {Auff\`eves}(2022)}]{Auffeves.22.PRXQ}%
  \BibitemOpen
  \bibfield  {author} {\bibinfo {author} {\bibfnamefont {A.}~\bibnamefont
  {Auff\`eves}},\ }\bibfield  {title} {\enquote {\bibinfo {title} {Quantum
  technologies need a quantum energy initiative},}\ }\href {\doibase
  10.1103/PRXQuantum.3.020101} {\bibfield  {journal} {\bibinfo  {journal} {PRX
  Quantum}\ }\textbf {\bibinfo {volume} {3}},\ \bibinfo {pages} {020101}
  (\bibinfo {year} {2022})}\BibitemShut {NoStop}%
\bibitem [{\citenamefont {Szilard}(1929)}]{Szilard.29.ZP}%
  \BibitemOpen
  \bibfield  {author} {\bibinfo {author} {\bibfnamefont {L.}~\bibnamefont
  {Szilard}},\ }\bibfield  {title} {\enquote {\bibinfo {title} {On the decrease
  in entropy in a thermodynamic system by the intervention of intelligent
  beings},}\ }\href {\doibase 10.1007/BF01341281} {\bibfield  {journal}
  {\bibinfo  {journal} {Z. Phys.}\ }\textbf {\bibinfo {volume} {53}},\ \bibinfo
  {pages} {840} (\bibinfo {year} {1929})}\BibitemShut {NoStop}%
\bibitem [{\citenamefont {Landauer}(1961)}]{Landauer.61.IBM}%
  \BibitemOpen
  \bibfield  {author} {\bibinfo {author} {\bibfnamefont {R.}~\bibnamefont
  {Landauer}},\ }\bibfield  {title} {\enquote {\bibinfo {title}
  {Irreversibility and {H}eat {G}eneration in the {C}omputing {P}rocess},}\
  }\href {\doibase 10.1147/rd.53.0183} {\bibfield  {journal} {\bibinfo
  {journal} {IBM J. Res. Dev.}\ }\textbf {\bibinfo {volume} {5}},\ \bibinfo
  {pages} {183} (\bibinfo {year} {1961})}\BibitemShut {NoStop}%
\bibitem [{\citenamefont {Maruyama}\ \emph {et~al.}(2009)\citenamefont
  {Maruyama}, \citenamefont {Nori},\ and\ \citenamefont
  {Vedral}}]{Maruyama.09.RMP}%
  \BibitemOpen
  \bibfield  {author} {\bibinfo {author} {\bibfnamefont {K.}~\bibnamefont
  {Maruyama}}, \bibinfo {author} {\bibfnamefont {F.}~\bibnamefont {Nori}}, \
  and\ \bibinfo {author} {\bibfnamefont {V.}~\bibnamefont {Vedral}},\
  }\bibfield  {title} {\enquote {\bibinfo {title} {Colloquium: The physics of
  maxwell's demon and information},}\ }\href {\doibase 10.1103/RevModPhys.81.1}
  {\bibfield  {journal} {\bibinfo  {journal} {Rev. Mod. Phys.}\ }\textbf
  {\bibinfo {volume} {81}},\ \bibinfo {pages} {1--23} (\bibinfo {year}
  {2009})}\BibitemShut {NoStop}%
\bibitem [{\citenamefont {Parrondo}\ \emph {et~al.}(2015)\citenamefont
  {Parrondo}, \citenamefont {Horowitz},\ and\ \citenamefont
  {Sagawa}}]{Parrondo.15.NP}%
  \BibitemOpen
  \bibfield  {author} {\bibinfo {author} {\bibfnamefont {J.}~\bibnamefont
  {Parrondo}}, \bibinfo {author} {\bibfnamefont {J.}~\bibnamefont {Horowitz}},
  \ and\ \bibinfo {author} {\bibfnamefont {T.}~\bibnamefont {Sagawa}},\
  }\bibfield  {title} {\enquote {\bibinfo {title} {Thermodynamics of
  information},}\ }\href {http://dx.doi.org/10.1038/nphys3230} {\bibfield
  {journal} {\bibinfo  {journal} {Nat. Phys.}\ }\textbf {\bibinfo {volume}
  {11}},\ \bibinfo {pages} {131} (\bibinfo {year} {2015})}\BibitemShut
  {NoStop}%
\bibitem [{\citenamefont {Goold}\ \emph {et~al.}(2016)\citenamefont {Goold},
  \citenamefont {Huber}, \citenamefont {Riera}, \citenamefont {del Rio},\ and\
  \citenamefont {Skrzypczyk}}]{Goold.16.JPA}%
  \BibitemOpen
  \bibfield  {author} {\bibinfo {author} {\bibfnamefont {J.}~\bibnamefont
  {Goold}}, \bibinfo {author} {\bibfnamefont {M.}~\bibnamefont {Huber}},
  \bibinfo {author} {\bibfnamefont {A.}~\bibnamefont {Riera}}, \bibinfo
  {author} {\bibfnamefont {L.}~\bibnamefont {del Rio}}, \ and\ \bibinfo
  {author} {\bibfnamefont {P.}~\bibnamefont {Skrzypczyk}},\ }\bibfield  {title}
  {\enquote {\bibinfo {title} {The role of quantum information in
  thermodynamics—a topical review},}\ }\href
  {http://stacks.iop.org/1751-8121/49/i=14/a=143001} {\bibfield  {journal}
  {\bibinfo  {journal} {J. Phys. A: Math. Theor.}\ }\textbf {\bibinfo {volume}
  {49}},\ \bibinfo {pages} {143001} (\bibinfo {year} {2016})}\BibitemShut
  {NoStop}%
\bibitem [{\citenamefont {Liu}\ \emph {et~al.}(2024)\citenamefont {Liu},
  \citenamefont {Lu}, \citenamefont {Wang},\ and\ \citenamefont
  {Jiang}}]{LiuJ.24.PRR}%
  \BibitemOpen
  \bibfield  {author} {\bibinfo {author} {\bibfnamefont {J.}~\bibnamefont
  {Liu}}, \bibinfo {author} {\bibfnamefont {J.}~\bibnamefont {Lu}}, \bibinfo
  {author} {\bibfnamefont {C.}~\bibnamefont {Wang}}, \ and\ \bibinfo {author}
  {\bibfnamefont {J.-H.}\ \bibnamefont {Jiang}},\ }\bibfield  {title} {\enquote
  {\bibinfo {title} {Inferring general links between energetics and information
  with unknown environment},}\ }\href {\doibase
  10.1103/PhysRevResearch.6.033202} {\bibfield  {journal} {\bibinfo  {journal}
  {Phys. Rev. Res.}\ }\textbf {\bibinfo {volume} {6}},\ \bibinfo {pages}
  {033202} (\bibinfo {year} {2024})}\BibitemShut {NoStop}%
\bibitem [{\citenamefont {Landi}\ and\ \citenamefont
  {Paternostro}(2021)}]{Landi.21.RMP}%
  \BibitemOpen
  \bibfield  {author} {\bibinfo {author} {\bibfnamefont {G.}~\bibnamefont
  {Landi}}\ and\ \bibinfo {author} {\bibfnamefont {M.}~\bibnamefont
  {Paternostro}},\ }\bibfield  {title} {\enquote {\bibinfo {title}
  {Irreversible entropy production: From classical to quantum},}\ }\href
  {\doibase 10.1103/RevModPhys.93.035008} {\bibfield  {journal} {\bibinfo
  {journal} {Rev. Mod. Phys.}\ }\textbf {\bibinfo {volume} {93}},\ \bibinfo
  {pages} {035008} (\bibinfo {year} {2021})}\BibitemShut {NoStop}%
\bibitem [{\citenamefont {Danageozian}\ \emph {et~al.}(2022)\citenamefont
  {Danageozian}, \citenamefont {Wilde},\ and\ \citenamefont
  {Buscemi}}]{Danageozian.22.PRXQ}%
  \BibitemOpen
  \bibfield  {author} {\bibinfo {author} {\bibfnamefont {A.}~\bibnamefont
  {Danageozian}}, \bibinfo {author} {\bibfnamefont {M.}~\bibnamefont {Wilde}},
  \ and\ \bibinfo {author} {\bibfnamefont {F.}~\bibnamefont {Buscemi}},\
  }\bibfield  {title} {\enquote {\bibinfo {title} {Thermodynamic constraints on
  quantum information gain and error correction: A triple trade-off},}\ }\href
  {\doibase 10.1103/PRXQuantum.3.020318} {\bibfield  {journal} {\bibinfo
  {journal} {PRX Quantum}\ }\textbf {\bibinfo {volume} {3}},\ \bibinfo {pages}
  {020318} (\bibinfo {year} {2022})}\BibitemShut {NoStop}%
\bibitem [{\citenamefont {Nielsen}\ \emph {et~al.}(1998)\citenamefont
  {Nielsen}, \citenamefont {Caves}, \citenamefont {Schumacher},\ and\
  \citenamefont {Barnum}}]{Nielsen.98.PRSL}%
  \BibitemOpen
  \bibfield  {author} {\bibinfo {author} {\bibfnamefont {M.}~\bibnamefont
  {Nielsen}}, \bibinfo {author} {\bibfnamefont {C.}~\bibnamefont {Caves}},
  \bibinfo {author} {\bibfnamefont {B.}~\bibnamefont {Schumacher}}, \ and\
  \bibinfo {author} {\bibfnamefont {H.}~\bibnamefont {Barnum}},\ }\bibfield
  {title} {\enquote {\bibinfo {title} {Information-theoretic approach to
  quantum error correction and reversible measurement},}\ }\href {\doibase
  10.1098/rspa.1998.0160} {\bibfield  {journal} {\bibinfo  {journal} {Proc. R.
  Soc. Lond.}\ }\textbf {\bibinfo {volume} {454}},\ \bibinfo {pages} {277}
  (\bibinfo {year} {1998})}\BibitemShut {NoStop}%
\bibitem [{\citenamefont {Sagawa}\ and\ \citenamefont
  {Ueda}(2009)}]{Sagawa.09.PRL}%
  \BibitemOpen
  \bibfield  {author} {\bibinfo {author} {\bibfnamefont {T.}~\bibnamefont
  {Sagawa}}\ and\ \bibinfo {author} {\bibfnamefont {M.}~\bibnamefont {Ueda}},\
  }\bibfield  {title} {\enquote {\bibinfo {title} {Minimal energy cost for
  thermodynamic information processing: Measurement and information erasure},}\
  }\href {\doibase 10.1103/PhysRevLett.102.250602} {\bibfield  {journal}
  {\bibinfo  {journal} {Phys. Rev. Lett.}\ }\textbf {\bibinfo {volume} {102}},\
  \bibinfo {pages} {250602} (\bibinfo {year} {2009})}\BibitemShut {NoStop}%
\bibitem [{\citenamefont {Horowitz}\ and\ \citenamefont
  {Jacobs}(2015)}]{Horowitz.15.PRL}%
  \BibitemOpen
  \bibfield  {author} {\bibinfo {author} {\bibfnamefont {J.}~\bibnamefont
  {Horowitz}}\ and\ \bibinfo {author} {\bibfnamefont {K.}~\bibnamefont
  {Jacobs}},\ }\bibfield  {title} {\enquote {\bibinfo {title} {Energy cost of
  controlling mesoscopic quantum systems},}\ }\href {\doibase
  10.1103/PhysRevLett.115.130501} {\bibfield  {journal} {\bibinfo  {journal}
  {Phys. Rev. Lett.}\ }\textbf {\bibinfo {volume} {115}},\ \bibinfo {pages}
  {130501} (\bibinfo {year} {2015})}\BibitemShut {NoStop}%
\bibitem [{\citenamefont {Boyd}\ \emph {et~al.}(2018)\citenamefont {Boyd},
  \citenamefont {Mandal},\ and\ \citenamefont {Crutchfield}}]{Boyd.18.PRX}%
  \BibitemOpen
  \bibfield  {author} {\bibinfo {author} {\bibfnamefont {A.}~\bibnamefont
  {Boyd}}, \bibinfo {author} {\bibfnamefont {D}~\bibnamefont {Mandal}}, \ and\
  \bibinfo {author} {\bibfnamefont {J.}~\bibnamefont {Crutchfield}},\
  }\bibfield  {title} {\enquote {\bibinfo {title} {Thermodynamics of
  modularity: Structural costs beyond the landauer bound},}\ }\href {\doibase
  10.1103/PhysRevX.8.031036} {\bibfield  {journal} {\bibinfo  {journal} {Phys.
  Rev. X}\ }\textbf {\bibinfo {volume} {8}},\ \bibinfo {pages} {031036}
  (\bibinfo {year} {2018})}\BibitemShut {NoStop}%
\bibitem [{\citenamefont {Abah}\ \emph {et~al.}(2019)\citenamefont {Abah},
  \citenamefont {Puebla}, \citenamefont {Kiely}, \citenamefont {Chiara},
  \citenamefont {Paternostro},\ and\ \citenamefont {Campbell}}]{Abah.19.NJP}%
  \BibitemOpen
  \bibfield  {author} {\bibinfo {author} {\bibfnamefont {O.}~\bibnamefont
  {Abah}}, \bibinfo {author} {\bibfnamefont {R.}~\bibnamefont {Puebla}},
  \bibinfo {author} {\bibfnamefont {A.}~\bibnamefont {Kiely}}, \bibinfo
  {author} {\bibfnamefont {G.~De}\ \bibnamefont {Chiara}}, \bibinfo {author}
  {\bibfnamefont {M.}~\bibnamefont {Paternostro}}, \ and\ \bibinfo {author}
  {\bibfnamefont {S.}~\bibnamefont {Campbell}},\ }\bibfield  {title} {\enquote
  {\bibinfo {title} {Energetic cost of quantum control protocols},}\ }\href
  {\doibase 10.1088/1367-2630/ab4c8c} {\bibfield  {journal} {\bibinfo
  {journal} {New J. Phys.}\ }\textbf {\bibinfo {volume} {21}},\ \bibinfo
  {pages} {103048} (\bibinfo {year} {2019})}\BibitemShut {NoStop}%
\bibitem [{\citenamefont {Deffner}(2021)}]{Deffner.21.EPL}%
  \BibitemOpen
  \bibfield  {author} {\bibinfo {author} {\bibfnamefont {S.}~\bibnamefont
  {Deffner}},\ }\bibfield  {title} {\enquote {\bibinfo {title} {Energetic cost
  of hamiltonian quantum gates},}\ }\href {\doibase
  10.1209/0295-5075/134/40002} {\bibfield  {journal} {\bibinfo  {journal}
  {Europhys. Lett.}\ }\textbf {\bibinfo {volume} {134}},\ \bibinfo {pages}
  {40002} (\bibinfo {year} {2021})}\BibitemShut {NoStop}%
\bibitem [{\citenamefont {Pearson}\ \emph {et~al.}(2021)\citenamefont
  {Pearson}, \citenamefont {Guryanova}, \citenamefont {Erker}, \citenamefont
  {Laird}, \citenamefont {Briggs}, \citenamefont {Huber},\ and\ \citenamefont
  {Ares}}]{Pearson.21.PRX}%
  \BibitemOpen
  \bibfield  {author} {\bibinfo {author} {\bibfnamefont {A.}~\bibnamefont
  {Pearson}}, \bibinfo {author} {\bibfnamefont {Y.}~\bibnamefont {Guryanova}},
  \bibinfo {author} {\bibfnamefont {P.}~\bibnamefont {Erker}}, \bibinfo
  {author} {\bibfnamefont {E.}~\bibnamefont {Laird}}, \bibinfo {author}
  {\bibfnamefont {G.}~\bibnamefont {Briggs}}, \bibinfo {author} {\bibfnamefont
  {M.}~\bibnamefont {Huber}}, \ and\ \bibinfo {author} {\bibfnamefont
  {N.}~\bibnamefont {Ares}},\ }\bibfield  {title} {\enquote {\bibinfo {title}
  {Measuring the thermodynamic cost of timekeeping},}\ }\href {\doibase
  10.1103/PhysRevX.11.021029} {\bibfield  {journal} {\bibinfo  {journal} {Phys.
  Rev. X}\ }\textbf {\bibinfo {volume} {11}},\ \bibinfo {pages} {021029}
  (\bibinfo {year} {2021})}\BibitemShut {NoStop}%
\bibitem [{\citenamefont {Chiribella}\ \emph {et~al.}(2022)\citenamefont
  {Chiribella}, \citenamefont {Meng}, \citenamefont {Renner},\ and\
  \citenamefont {Yung}}]{Chiribella.22.NC}%
  \BibitemOpen
  \bibfield  {author} {\bibinfo {author} {\bibfnamefont {G.}~\bibnamefont
  {Chiribella}}, \bibinfo {author} {\bibfnamefont {F.}~\bibnamefont {Meng}},
  \bibinfo {author} {\bibfnamefont {R.}~\bibnamefont {Renner}}, \ and\ \bibinfo
  {author} {\bibfnamefont {M.}~\bibnamefont {Yung}},\ }\bibfield  {title}
  {\enquote {\bibinfo {title} {The nonequilibrium cost of accurate information
  processing},}\ }\href {\doibase 10.1038/s41467-022-34541-w} {\bibfield
  {journal} {\bibinfo  {journal} {Nat. Commun.}\ }\textbf {\bibinfo {volume}
  {13}},\ \bibinfo {pages} {7155} (\bibinfo {year} {2022})}\BibitemShut
  {NoStop}%
\bibitem [{\citenamefont {Woods}\ and\ \citenamefont
  {Horodecki}(2023)}]{Woods.23.PRX}%
  \BibitemOpen
  \bibfield  {author} {\bibinfo {author} {\bibfnamefont {M.}~\bibnamefont
  {Woods}}\ and\ \bibinfo {author} {\bibfnamefont {M.}~\bibnamefont
  {Horodecki}},\ }\bibfield  {title} {\enquote {\bibinfo {title} {Autonomous
  quantum devices: When are they realizable without additional thermodynamic
  costs?}}\ }\href {\doibase 10.1103/PhysRevX.13.011016} {\bibfield  {journal}
  {\bibinfo  {journal} {Phys. Rev. X}\ }\textbf {\bibinfo {volume} {13}},\
  \bibinfo {pages} {011016} (\bibinfo {year} {2023})}\BibitemShut {NoStop}%
\bibitem [{\citenamefont {Bloch}\ \emph {et~al.}(2008)\citenamefont {Bloch},
  \citenamefont {Dalibard},\ and\ \citenamefont {Zwerger}}]{Bloch.08.RMP}%
  \BibitemOpen
  \bibfield  {author} {\bibinfo {author} {\bibfnamefont {I.}~\bibnamefont
  {Bloch}}, \bibinfo {author} {\bibfnamefont {J.}~\bibnamefont {Dalibard}}, \
  and\ \bibinfo {author} {\bibfnamefont {W.}~\bibnamefont {Zwerger}},\
  }\bibfield  {title} {\enquote {\bibinfo {title} {Many-body physics with
  ultracold gases},}\ }\href {\doibase 10.1103/RevModPhys.80.885} {\bibfield
  {journal} {\bibinfo  {journal} {Rev. Mod. Phys.}\ }\textbf {\bibinfo {volume}
  {80}},\ \bibinfo {pages} {885} (\bibinfo {year} {2008})}\BibitemShut
  {NoStop}%
\bibitem [{\citenamefont {Chin}\ \emph {et~al.}(2012)\citenamefont {Chin},
  \citenamefont {Huelga},\ and\ \citenamefont {Plenio}}]{Chin.12.PRL}%
  \BibitemOpen
  \bibfield  {author} {\bibinfo {author} {\bibfnamefont {A.}~\bibnamefont
  {Chin}}, \bibinfo {author} {\bibfnamefont {S.}~\bibnamefont {Huelga}}, \ and\
  \bibinfo {author} {\bibfnamefont {M.}~\bibnamefont {Plenio}},\ }\bibfield
  {title} {\enquote {\bibinfo {title} {Quantum metrology in non-markovian
  environments},}\ }\href {\doibase 10.1103/PhysRevLett.109.233601} {\bibfield
  {journal} {\bibinfo  {journal} {Phys. Rev. Lett.}\ }\textbf {\bibinfo
  {volume} {109}},\ \bibinfo {pages} {233601} (\bibinfo {year}
  {2012})}\BibitemShut {NoStop}%
\bibitem [{\citenamefont {Georgescu}\ \emph {et~al.}(2014)\citenamefont
  {Georgescu}, \citenamefont {Ashhab},\ and\ \citenamefont
  {Nori}}]{Georgescu.14.RMP}%
  \BibitemOpen
  \bibfield  {author} {\bibinfo {author} {\bibfnamefont {I.}~\bibnamefont
  {Georgescu}}, \bibinfo {author} {\bibfnamefont {S.}~\bibnamefont {Ashhab}}, \
  and\ \bibinfo {author} {\bibfnamefont {F.}~\bibnamefont {Nori}},\ }\bibfield
  {title} {\enquote {\bibinfo {title} {Quantum simulation},}\ }\href {\doibase
  10.1103/RevModPhys.86.153} {\bibfield  {journal} {\bibinfo  {journal} {Rev.
  Mod. Phys.}\ }\textbf {\bibinfo {volume} {86}},\ \bibinfo {pages} {153--185}
  (\bibinfo {year} {2014})}\BibitemShut {NoStop}%
\bibitem [{\citenamefont {Adesso}\ \emph {et~al.}(2016)\citenamefont {Adesso},
  \citenamefont {Bromley},\ and\ \citenamefont {Cianciaruso}}]{Adesso.16.JPA}%
  \BibitemOpen
  \bibfield  {author} {\bibinfo {author} {\bibfnamefont {G.}~\bibnamefont
  {Adesso}}, \bibinfo {author} {\bibfnamefont {T.}~\bibnamefont {Bromley}}, \
  and\ \bibinfo {author} {\bibfnamefont {M.}~\bibnamefont {Cianciaruso}},\
  }\bibfield  {title} {\enquote {\bibinfo {title} {Measures and applications of
  quantum correlations},}\ }\href {\doibase 10.1088/1751-8113/49/47/473001}
  {\bibfield  {journal} {\bibinfo  {journal} {J. Phys. A: Math. Theor.}\
  }\textbf {\bibinfo {volume} {49}},\ \bibinfo {pages} {473001} (\bibinfo
  {year} {2016})}\BibitemShut {NoStop}%
\bibitem [{\citenamefont {Koch}(2016)}]{Koch.16.JPA}%
  \BibitemOpen
  \bibfield  {author} {\bibinfo {author} {\bibfnamefont {C.}~\bibnamefont
  {Koch}},\ }\bibfield  {title} {\enquote {\bibinfo {title} {Controlling open
  quantum systems: tools, achievements, and limitations},}\ }\href {\doibase
  10.1088/0953-8984/28/21/213001} {\bibfield  {journal} {\bibinfo  {journal}
  {J. Phys.: Condensed Matter}\ }\textbf {\bibinfo {volume} {28}},\ \bibinfo
  {pages} {213001} (\bibinfo {year} {2016})}\BibitemShut {NoStop}%
\bibitem [{\citenamefont {Chiara}\ and\ \citenamefont
  {Sanpera}(2018)}]{Chiara.18.RPP}%
  \BibitemOpen
  \bibfield  {author} {\bibinfo {author} {\bibfnamefont {G.}~\bibnamefont
  {Chiara}}\ and\ \bibinfo {author} {\bibfnamefont {A.}~\bibnamefont
  {Sanpera}},\ }\bibfield  {title} {\enquote {\bibinfo {title} {Genuine quantum
  correlations in quantum many-body systems: a review of recent progress},}\
  }\href {\doibase 10.1088/1361-6633/aabf61} {\bibfield  {journal} {\bibinfo
  {journal} {Rep. Prog. Phys.}\ }\textbf {\bibinfo {volume} {81}},\ \bibinfo
  {pages} {074002} (\bibinfo {year} {2018})}\BibitemShut {NoStop}%
\bibitem [{\citenamefont {Shahandeh}\ \emph {et~al.}(2019)\citenamefont
  {Shahandeh}, \citenamefont {Lund},\ and\ \citenamefont
  {Ralph}}]{Shahandeh.19.PRA}%
  \BibitemOpen
  \bibfield  {author} {\bibinfo {author} {\bibfnamefont {F.}~\bibnamefont
  {Shahandeh}}, \bibinfo {author} {\bibfnamefont {A.}~\bibnamefont {Lund}}, \
  and\ \bibinfo {author} {\bibfnamefont {T.}~\bibnamefont {Ralph}},\ }\bibfield
   {title} {\enquote {\bibinfo {title} {Quantum correlations and global
  coherence in distributed quantum computing},}\ }\href {\doibase
  10.1103/PhysRevA.99.052303} {\bibfield  {journal} {\bibinfo  {journal} {Phys.
  Rev. A}\ }\textbf {\bibinfo {volume} {99}},\ \bibinfo {pages} {052303}
  (\bibinfo {year} {2019})}\BibitemShut {NoStop}%
\bibitem [{\citenamefont {Head-Marsden}\ \emph {et~al.}(2021)\citenamefont
  {Head-Marsden}, \citenamefont {Flick}, \citenamefont {Ciccarino},\ and\
  \citenamefont {Narang}}]{Narang.20.CR}%
  \BibitemOpen
  \bibfield  {author} {\bibinfo {author} {\bibfnamefont {K.}~\bibnamefont
  {Head-Marsden}}, \bibinfo {author} {\bibfnamefont {J.}~\bibnamefont {Flick}},
  \bibinfo {author} {\bibfnamefont {C.}~\bibnamefont {Ciccarino}}, \ and\
  \bibinfo {author} {\bibfnamefont {P.}~\bibnamefont {Narang}},\ }\bibfield
  {title} {\enquote {\bibinfo {title} {Quantum information and algorithms for
  correlated quantum matter},}\ }\href {\doibase 10.1021/acs.chemrev.0c00620}
  {\bibfield  {journal} {\bibinfo  {journal} {Chem. Rev.}\ }\textbf {\bibinfo
  {volume} {121}},\ \bibinfo {pages} {3061} (\bibinfo {year}
  {2021})}\BibitemShut {NoStop}%
\bibitem [{\citenamefont {Britton}\ \emph {et~al.}(2012)\citenamefont
  {Britton}, \citenamefont {Sawyer}, \citenamefont {Keith}, \citenamefont
  {Wang}, \citenamefont {Freericks}, \citenamefont {Uys}, \citenamefont
  {Biercuk},\ and\ \citenamefont {Bollinger}}]{Britton.12.N}%
  \BibitemOpen
  \bibfield  {author} {\bibinfo {author} {\bibfnamefont {J.}~\bibnamefont
  {Britton}}, \bibinfo {author} {\bibfnamefont {B.}~\bibnamefont {Sawyer}},
  \bibinfo {author} {\bibfnamefont {A.}~\bibnamefont {Keith}}, \bibinfo
  {author} {\bibfnamefont {C.}~\bibnamefont {Wang}}, \bibinfo {author}
  {\bibfnamefont {J.}~\bibnamefont {Freericks}}, \bibinfo {author}
  {\bibfnamefont {H.}~\bibnamefont {Uys}}, \bibinfo {author} {\bibfnamefont
  {M.}~\bibnamefont {Biercuk}}, \ and\ \bibinfo {author} {\bibfnamefont
  {J.}~\bibnamefont {Bollinger}},\ }\bibfield  {title} {\enquote {\bibinfo
  {title} {Engineered two-dimensional ising interactions in a trapped-ion
  quantum simulator with hundreds of spins},}\ }\href {\doibase
  10.1038/nature10981} {\bibfield  {journal} {\bibinfo  {journal} {Nature}\
  }\textbf {\bibinfo {volume} {484}},\ \bibinfo {pages} {489} (\bibinfo {year}
  {2012})}\BibitemShut {NoStop}%
\bibitem [{\citenamefont {Labuhn}\ \emph {et~al.}(2016)\citenamefont {Labuhn},
  \citenamefont {Barredo}, \citenamefont {Ravets}, \citenamefont
  {de~L\'es\'eleuc}, \citenamefont {Macrì}, \citenamefont {Lahaye},\ and\
  \citenamefont {Browaeys}}]{Labuhn.16.N}%
  \BibitemOpen
  \bibfield  {author} {\bibinfo {author} {\bibfnamefont {H.}~\bibnamefont
  {Labuhn}}, \bibinfo {author} {\bibfnamefont {D.}~\bibnamefont {Barredo}},
  \bibinfo {author} {\bibfnamefont {S.}~\bibnamefont {Ravets}}, \bibinfo
  {author} {\bibfnamefont {S.}~\bibnamefont {de~L\'es\'eleuc}}, \bibinfo
  {author} {\bibfnamefont {T.}~\bibnamefont {Macrì}}, \bibinfo {author}
  {\bibfnamefont {T.}~\bibnamefont {Lahaye}}, \ and\ \bibinfo {author}
  {\bibfnamefont {A.}~\bibnamefont {Browaeys}},\ }\bibfield  {title} {\enquote
  {\bibinfo {title} {Tunable two-dimensional arrays of single rydberg atoms for
  realizing quantum ising models},}\ }\href
  {http://dx.doi.org/10.1038/nature18274} {\bibfield  {journal} {\bibinfo
  {journal} {Nature}\ }\textbf {\bibinfo {volume} {534}},\ \bibinfo {pages}
  {667} (\bibinfo {year} {2016})}\BibitemShut {NoStop}%
\bibitem [{\citenamefont {Monroe}\ \emph {et~al.}(2021)\citenamefont {Monroe},
  \citenamefont {Campbell}, \citenamefont {Duan}, \citenamefont {Gong},
  \citenamefont {Gorshkov}, \citenamefont {Hess}, \citenamefont {Islam},
  \citenamefont {Kim}, \citenamefont {Linke}, \citenamefont {Pagano},
  \citenamefont {Richerme}, \citenamefont {Senko},\ and\ \citenamefont
  {Yao}}]{Monroe.21.RMP}%
  \BibitemOpen
  \bibfield  {author} {\bibinfo {author} {\bibfnamefont {C.}~\bibnamefont
  {Monroe}}, \bibinfo {author} {\bibfnamefont {W.~C.}\ \bibnamefont
  {Campbell}}, \bibinfo {author} {\bibfnamefont {L.-M.}\ \bibnamefont {Duan}},
  \bibinfo {author} {\bibfnamefont {Z.-X.}\ \bibnamefont {Gong}}, \bibinfo
  {author} {\bibfnamefont {A.~V.}\ \bibnamefont {Gorshkov}}, \bibinfo {author}
  {\bibfnamefont {P.~W.}\ \bibnamefont {Hess}}, \bibinfo {author}
  {\bibfnamefont {R.}~\bibnamefont {Islam}}, \bibinfo {author} {\bibfnamefont
  {K.}~\bibnamefont {Kim}}, \bibinfo {author} {\bibfnamefont {N.~M.}\
  \bibnamefont {Linke}}, \bibinfo {author} {\bibfnamefont {G.}~\bibnamefont
  {Pagano}}, \bibinfo {author} {\bibfnamefont {P.}~\bibnamefont {Richerme}},
  \bibinfo {author} {\bibfnamefont {C.}~\bibnamefont {Senko}}, \ and\ \bibinfo
  {author} {\bibfnamefont {N.}~\bibnamefont {Yao}},\ }\bibfield  {title}
  {\enquote {\bibinfo {title} {Programmable quantum simulations of spin systems
  with trapped ions},}\ }\href {\doibase 10.1103/RevModPhys.93.025001}
  {\bibfield  {journal} {\bibinfo  {journal} {Rev. Mod. Phys.}\ }\textbf
  {\bibinfo {volume} {93}},\ \bibinfo {pages} {025001} (\bibinfo {year}
  {2021})}\BibitemShut {NoStop}%
\bibitem [{\citenamefont {Periwal}\ \emph {et~al.}(2021)\citenamefont
  {Periwal}, \citenamefont {Cooper}, \citenamefont {Kunkel}, \citenamefont
  {Wienand}, \citenamefont {Davis},\ and\ \citenamefont
  {Schleier-Smith}}]{Periwal.21.N}%
  \BibitemOpen
  \bibfield  {author} {\bibinfo {author} {\bibfnamefont {A.}~\bibnamefont
  {Periwal}}, \bibinfo {author} {\bibfnamefont {E.}~\bibnamefont {Cooper}},
  \bibinfo {author} {\bibfnamefont {P.}~\bibnamefont {Kunkel}}, \bibinfo
  {author} {\bibfnamefont {J.}~\bibnamefont {Wienand}}, \bibinfo {author}
  {\bibfnamefont {E.}~\bibnamefont {Davis}}, \ and\ \bibinfo {author}
  {\bibfnamefont {M.}~\bibnamefont {Schleier-Smith}},\ }\bibfield  {title}
  {\enquote {\bibinfo {title} {Programmable interactions and emergent geometry
  in an array of atom clouds},}\ }\href {\doibase 10.1038/s41586-021-04156-0}
  {\bibfield  {journal} {\bibinfo  {journal} {Nature}\ }\textbf {\bibinfo
  {volume} {600}},\ \bibinfo {pages} {630} (\bibinfo {year}
  {2021})}\BibitemShut {NoStop}%
\bibitem [{\citenamefont {Mivehvar}\ \emph {et~al.}(2021)\citenamefont
  {Mivehvar}, \citenamefont {Piazza}, \citenamefont {Donner},\ and\
  \citenamefont {Ritsch}}]{Mivehvar.21.AP}%
  \BibitemOpen
  \bibfield  {author} {\bibinfo {author} {\bibfnamefont {F.}~\bibnamefont
  {Mivehvar}}, \bibinfo {author} {\bibfnamefont {F.}~\bibnamefont {Piazza}},
  \bibinfo {author} {\bibfnamefont {T.}~\bibnamefont {Donner}}, \ and\ \bibinfo
  {author} {\bibfnamefont {H.}~\bibnamefont {Ritsch}},\ }\bibfield  {title}
  {\enquote {\bibinfo {title} {Cavity {QED} with quantum gases: new paradigms
  in many-body physics},}\ }\href {\doibase 10.1080/00018732.2021.1969727}
  {\bibfield  {journal} {\bibinfo  {journal} {Adv. Phys.}\ }\textbf {\bibinfo
  {volume} {70}},\ \bibinfo {pages} {1} (\bibinfo {year} {2021})}\BibitemShut
  {NoStop}%
\bibitem [{\citenamefont {Zwolak}\ and\ \citenamefont
  {Taylor}(2023)}]{Zwolak.23.RMP}%
  \BibitemOpen
  \bibfield  {author} {\bibinfo {author} {\bibfnamefont {J.}~\bibnamefont
  {Zwolak}}\ and\ \bibinfo {author} {\bibfnamefont {J.}~\bibnamefont
  {Taylor}},\ }\bibfield  {title} {\enquote {\bibinfo {title} {Colloquium:
  Advances in automation of quantum dot devices control},}\ }\href {\doibase
  10.1103/RevModPhys.95.011006} {\bibfield  {journal} {\bibinfo  {journal}
  {Rev. Mod. Phys.}\ }\textbf {\bibinfo {volume} {95}},\ \bibinfo {pages}
  {011006} (\bibinfo {year} {2023})}\BibitemShut {NoStop}%
\bibitem [{\citenamefont {Mi}\ \emph {et~al.}(2024)\citenamefont {Mi},
  \citenamefont {Michailidis}, \citenamefont {Shabani}, \citenamefont {Miao},
  \citenamefont {Klimov}, \citenamefont {Lloyd}, \citenamefont {Rosenberg},
  \citenamefont {Acharya}, \citenamefont {Aleiner}, \citenamefont {Andersen},\
  and\ \citenamefont {et. al.}}]{MiX.24.S}%
  \BibitemOpen
  \bibfield  {author} {\bibinfo {author} {\bibfnamefont {X.}~\bibnamefont
  {Mi}}, \bibinfo {author} {\bibfnamefont {A.}~\bibnamefont {Michailidis}},
  \bibinfo {author} {\bibfnamefont {S.}~\bibnamefont {Shabani}}, \bibinfo
  {author} {\bibfnamefont {K.}~\bibnamefont {Miao}}, \bibinfo {author}
  {\bibfnamefont {P.}~\bibnamefont {Klimov}}, \bibinfo {author} {\bibfnamefont
  {J.}~\bibnamefont {Lloyd}}, \bibinfo {author} {\bibfnamefont
  {E.}~\bibnamefont {Rosenberg}}, \bibinfo {author} {\bibfnamefont
  {R.}~\bibnamefont {Acharya}}, \bibinfo {author} {\bibfnamefont
  {I.}~\bibnamefont {Aleiner}}, \bibinfo {author} {\bibfnamefont {T.~I.}\
  \bibnamefont {Andersen}}, \ and\ \bibinfo {author} {\bibnamefont {et. al.}},\
  }\bibfield  {title} {\enquote {\bibinfo {title} {Stable quantum-correlated
  many-body states through engineered dissipation},}\ }\href {\doibase
  10.1126/science.adh9932} {\bibfield  {journal} {\bibinfo  {journal}
  {Science}\ }\textbf {\bibinfo {volume} {383}},\ \bibinfo {pages} {1332}
  (\bibinfo {year} {2024})}\BibitemShut {NoStop}%
\bibitem [{\citenamefont {Defenu}\ \emph {et~al.}(2023)\citenamefont {Defenu},
  \citenamefont {Donner}, \citenamefont {Macr\`{\i}}, \citenamefont {Pagano},
  \citenamefont {Ruffo},\ and\ \citenamefont {Trombettoni}}]{Defenu.23.RMP}%
  \BibitemOpen
  \bibfield  {author} {\bibinfo {author} {\bibfnamefont {N.}~\bibnamefont
  {Defenu}}, \bibinfo {author} {\bibfnamefont {T.}~\bibnamefont {Donner}},
  \bibinfo {author} {\bibfnamefont {T.}~\bibnamefont {Macr\`{\i}}}, \bibinfo
  {author} {\bibfnamefont {G.}~\bibnamefont {Pagano}}, \bibinfo {author}
  {\bibfnamefont {S.}~\bibnamefont {Ruffo}}, \ and\ \bibinfo {author}
  {\bibfnamefont {A.}~\bibnamefont {Trombettoni}},\ }\bibfield  {title}
  {\enquote {\bibinfo {title} {Long-range interacting quantum systems},}\
  }\href {\doibase 10.1103/RevModPhys.95.035002} {\bibfield  {journal}
  {\bibinfo  {journal} {Rev. Mod. Phys.}\ }\textbf {\bibinfo {volume} {95}},\
  \bibinfo {pages} {035002} (\bibinfo {year} {2023})}\BibitemShut {NoStop}%
\bibitem [{\citenamefont {Feng}\ \emph {et~al.}(2023)\citenamefont {Feng},
  \citenamefont {Katz}, \citenamefont {Haack}, \citenamefont {Maghrebi},
  \citenamefont {Gorshkov}, \citenamefont {Gong}, \citenamefont {Cetina},\ and\
  \citenamefont {Monroe}}]{Feng.23.N}%
  \BibitemOpen
  \bibfield  {author} {\bibinfo {author} {\bibfnamefont {L.}~\bibnamefont
  {Feng}}, \bibinfo {author} {\bibfnamefont {O.}~\bibnamefont {Katz}}, \bibinfo
  {author} {\bibfnamefont {C.}~\bibnamefont {Haack}}, \bibinfo {author}
  {\bibfnamefont {M.}~\bibnamefont {Maghrebi}}, \bibinfo {author}
  {\bibfnamefont {A.}~\bibnamefont {Gorshkov}}, \bibinfo {author}
  {\bibfnamefont {Z.}~\bibnamefont {Gong}}, \bibinfo {author} {\bibfnamefont
  {M.}~\bibnamefont {Cetina}}, \ and\ \bibinfo {author} {\bibfnamefont
  {C.}~\bibnamefont {Monroe}},\ }\bibfield  {title} {\enquote {\bibinfo {title}
  {Continuous symmetry breaking in a trapped-ion spin chain},}\ }\href
  {\doibase 10.1038/s41586-023-06656-7} {\bibfield  {journal} {\bibinfo
  {journal} {Nature}\ }\textbf {\bibinfo {volume} {623}},\ \bibinfo {pages}
  {713} (\bibinfo {year} {2023})}\BibitemShut {NoStop}%
\bibitem [{\citenamefont {Sagawa}\ and\ \citenamefont
  {Ueda}(2012)}]{Sagawa.12.PRL}%
  \BibitemOpen
  \bibfield  {author} {\bibinfo {author} {\bibfnamefont {T.}~\bibnamefont
  {Sagawa}}\ and\ \bibinfo {author} {\bibfnamefont {M.}~\bibnamefont {Ueda}},\
  }\bibfield  {title} {\enquote {\bibinfo {title} {Fluctuation theorem with
  information exchange: Role of correlations in stochastic thermodynamics},}\
  }\href {\doibase 10.1103/PhysRevLett.109.180602} {\bibfield  {journal}
  {\bibinfo  {journal} {Phys. Rev. Lett.}\ }\textbf {\bibinfo {volume} {109}},\
  \bibinfo {pages} {180602} (\bibinfo {year} {2012})}\BibitemShut {NoStop}%
\bibitem [{\citenamefont {Bera}\ \emph {et~al.}(2017)\citenamefont {Bera},
  \citenamefont {Riera}, \citenamefont {Lewenstein},\ and\ \citenamefont
  {Winter}}]{Bera.17.NC}%
  \BibitemOpen
  \bibfield  {author} {\bibinfo {author} {\bibfnamefont {M.}~\bibnamefont
  {Bera}}, \bibinfo {author} {\bibfnamefont {A.}~\bibnamefont {Riera}},
  \bibinfo {author} {\bibfnamefont {M.}~\bibnamefont {Lewenstein}}, \ and\
  \bibinfo {author} {\bibfnamefont {A.}~\bibnamefont {Winter}},\ }\bibfield
  {title} {\enquote {\bibinfo {title} {Generalized laws of thermodynamics in
  the presence of correlations},}\ }\href {\doibase 10.1038/s41467-017-02370-x}
  {\bibfield  {journal} {\bibinfo  {journal} {Nat. Commun.}\ }\textbf {\bibinfo
  {volume} {8}},\ \bibinfo {pages} {2180} (\bibinfo {year} {2017})}\BibitemShut
  {NoStop}%
\bibitem [{\citenamefont {Sapienza}\ \emph {et~al.}(2019)\citenamefont
  {Sapienza}, \citenamefont {Cerisola},\ and\ \citenamefont
  {Roncaglia}}]{Sapienza.19.NC}%
  \BibitemOpen
  \bibfield  {author} {\bibinfo {author} {\bibfnamefont {F.}~\bibnamefont
  {Sapienza}}, \bibinfo {author} {\bibfnamefont {F.}~\bibnamefont {Cerisola}},
  \ and\ \bibinfo {author} {\bibfnamefont {A.}~\bibnamefont {Roncaglia}},\
  }\bibfield  {title} {\enquote {\bibinfo {title} {Correlations as a resource
  in quantum thermodynamics},}\ }\href {\doibase 10.1038/s41467-019-10572-8}
  {\bibfield  {journal} {\bibinfo  {journal} {Nat. Commun.}\ }\textbf {\bibinfo
  {volume} {10}},\ \bibinfo {pages} {2492} (\bibinfo {year}
  {2019})}\BibitemShut {NoStop}%
\bibitem [{\citenamefont {Oppenheim}\ \emph {et~al.}(2002)\citenamefont
  {Oppenheim}, \citenamefont {Horodecki}, \citenamefont {Horodecki},\ and\
  \citenamefont {Horodecki}}]{Oppenheim.02.PRL}%
  \BibitemOpen
  \bibfield  {author} {\bibinfo {author} {\bibfnamefont {J.}~\bibnamefont
  {Oppenheim}}, \bibinfo {author} {\bibfnamefont {M.}~\bibnamefont
  {Horodecki}}, \bibinfo {author} {\bibfnamefont {P.}~\bibnamefont
  {Horodecki}}, \ and\ \bibinfo {author} {\bibfnamefont {R.}~\bibnamefont
  {Horodecki}},\ }\bibfield  {title} {\enquote {\bibinfo {title}
  {Thermodynamical approach to quantifying quantum correlations},}\ }\href
  {\doibase 10.1103/PhysRevLett.89.180402} {\bibfield  {journal} {\bibinfo
  {journal} {Phys. Rev. Lett.}\ }\textbf {\bibinfo {volume} {89}},\ \bibinfo
  {pages} {180402} (\bibinfo {year} {2002})}\BibitemShut {NoStop}%
\bibitem [{\citenamefont {Perarnau-Llobet}\ \emph {et~al.}(2015)\citenamefont
  {Perarnau-Llobet}, \citenamefont {Hovhannisyan}, \citenamefont {Huber},
  \citenamefont {Skrzypczyk}, \citenamefont {Brunner},\ and\ \citenamefont
  {Ac\'{\i}n}}]{Llobet.15.PRX}%
  \BibitemOpen
  \bibfield  {author} {\bibinfo {author} {\bibfnamefont {M.}~\bibnamefont
  {Perarnau-Llobet}}, \bibinfo {author} {\bibfnamefont {K.}~\bibnamefont
  {Hovhannisyan}}, \bibinfo {author} {\bibfnamefont {M.}~\bibnamefont {Huber}},
  \bibinfo {author} {\bibfnamefont {P.}~\bibnamefont {Skrzypczyk}}, \bibinfo
  {author} {\bibfnamefont {N.}~\bibnamefont {Brunner}}, \ and\ \bibinfo
  {author} {\bibfnamefont {A.}~\bibnamefont {Ac\'{\i}n}},\ }\bibfield  {title}
  {\enquote {\bibinfo {title} {Extractable work from correlations},}\ }\href
  {\doibase 10.1103/PhysRevX.5.041011} {\bibfield  {journal} {\bibinfo
  {journal} {Phys. Rev. X}\ }\textbf {\bibinfo {volume} {5}},\ \bibinfo {pages}
  {041011} (\bibinfo {year} {2015})}\BibitemShut {NoStop}%
\bibitem [{\citenamefont {Manzano}\ \emph {et~al.}(2018)\citenamefont
  {Manzano}, \citenamefont {Plastina},\ and\ \citenamefont
  {Zambrini}}]{Manzano.18.PRL}%
  \BibitemOpen
  \bibfield  {author} {\bibinfo {author} {\bibfnamefont {G.}~\bibnamefont
  {Manzano}}, \bibinfo {author} {\bibfnamefont {F.}~\bibnamefont {Plastina}}, \
  and\ \bibinfo {author} {\bibfnamefont {R.}~\bibnamefont {Zambrini}},\
  }\bibfield  {title} {\enquote {\bibinfo {title} {Optimal work extraction and
  thermodynamics of quantum measurements and correlations},}\ }\href {\doibase
  10.1103/PhysRevLett.121.120602} {\bibfield  {journal} {\bibinfo  {journal}
  {Phys. Rev. Lett.}\ }\textbf {\bibinfo {volume} {121}},\ \bibinfo {pages}
  {120602} (\bibinfo {year} {2018})}\BibitemShut {NoStop}%
\bibitem [{\citenamefont {Micadei}\ \emph {et~al.}(2019)\citenamefont
  {Micadei}, \citenamefont {Peterson}, \citenamefont {Souza}, \citenamefont
  {Sarthour}, \citenamefont {Oliveira}, \citenamefont {Landi}, \citenamefont
  {Batalh{\~a}o}, \citenamefont {Serra},\ and\ \citenamefont
  {Lutz}}]{Micadei.19.NC}%
  \BibitemOpen
  \bibfield  {author} {\bibinfo {author} {\bibfnamefont {K.}~\bibnamefont
  {Micadei}}, \bibinfo {author} {\bibfnamefont {J.}~\bibnamefont {Peterson}},
  \bibinfo {author} {\bibfnamefont {A.}~\bibnamefont {Souza}}, \bibinfo
  {author} {\bibfnamefont {R.}~\bibnamefont {Sarthour}}, \bibinfo {author}
  {\bibfnamefont {I.}~\bibnamefont {Oliveira}}, \bibinfo {author}
  {\bibfnamefont {G.}~\bibnamefont {Landi}}, \bibinfo {author} {\bibfnamefont
  {T.}~\bibnamefont {Batalh{\~a}o}}, \bibinfo {author} {\bibfnamefont
  {R.}~\bibnamefont {Serra}}, \ and\ \bibinfo {author} {\bibfnamefont
  {E.}~\bibnamefont {Lutz}},\ }\bibfield  {title} {\enquote {\bibinfo {title}
  {Reversing the direction of heat flow using quantum correlations},}\ }\href
  {\doibase 10.1038/s41467-019-10333-7} {\bibfield  {journal} {\bibinfo
  {journal} {Nat. Commun.}\ }\textbf {\bibinfo {volume} {10}},\ \bibinfo
  {pages} {2456} (\bibinfo {year} {2019})}\BibitemShut {NoStop}%
\bibitem [{\citenamefont {Lipka-Bartosik}\ \emph {et~al.}(2024)\citenamefont
  {Lipka-Bartosik}, \citenamefont {Diotallevi},\ and\ \citenamefont
  {Bakhshinezhad}}]{Bartosik.24.PRL}%
  \BibitemOpen
  \bibfield  {author} {\bibinfo {author} {\bibfnamefont {P.}~\bibnamefont
  {Lipka-Bartosik}}, \bibinfo {author} {\bibfnamefont {G.}~\bibnamefont
  {Diotallevi}}, \ and\ \bibinfo {author} {\bibfnamefont {P.}~\bibnamefont
  {Bakhshinezhad}},\ }\bibfield  {title} {\enquote {\bibinfo {title}
  {Fundamental limits on anomalous energy flows in correlated quantum
  systems},}\ }\href {\doibase 10.1103/PhysRevLett.132.140402} {\bibfield
  {journal} {\bibinfo  {journal} {Phys. Rev. Lett.}\ }\textbf {\bibinfo
  {volume} {132}},\ \bibinfo {pages} {140402} (\bibinfo {year}
  {2024})}\BibitemShut {NoStop}%
\bibitem [{\citenamefont {Huber}\ \emph {et~al.}(2015)\citenamefont {Huber},
  \citenamefont {Perarnau-Llobet}, \citenamefont {Hovhannisyan}, \citenamefont
  {Skrzypczyk}, \citenamefont {Kl\"ockl}, \citenamefont {Brunner},\ and\
  \citenamefont {Ac\'in}}]{Huber.15.NJP}%
  \BibitemOpen
  \bibfield  {author} {\bibinfo {author} {\bibfnamefont {M.}~\bibnamefont
  {Huber}}, \bibinfo {author} {\bibfnamefont {M.}~\bibnamefont
  {Perarnau-Llobet}}, \bibinfo {author} {\bibfnamefont {K.}~\bibnamefont
  {Hovhannisyan}}, \bibinfo {author} {\bibfnamefont {P.}~\bibnamefont
  {Skrzypczyk}}, \bibinfo {author} {\bibfnamefont {C.}~\bibnamefont
  {Kl\"ockl}}, \bibinfo {author} {\bibfnamefont {N.}~\bibnamefont {Brunner}}, \
  and\ \bibinfo {author} {\bibfnamefont {A.}~\bibnamefont {Ac\'in}},\
  }\bibfield  {title} {\enquote {\bibinfo {title} {Thermodynamic cost of
  creating correlations},}\ }\href {\doibase 10.1088/1367-2630/17/6/065008}
  {\bibfield  {journal} {\bibinfo  {journal} {New J. Phys.}\ }\textbf {\bibinfo
  {volume} {17}},\ \bibinfo {pages} {065008} (\bibinfo {year}
  {2015})}\BibitemShut {NoStop}%
\bibitem [{\citenamefont {Rio}\ \emph {et~al.}(2011)\citenamefont {Rio},
  \citenamefont {Åberg}, \citenamefont {Renner}, \citenamefont {Dahlsten},\
  and\ \citenamefont {Vedral}}]{Rio.11.N}%
  \BibitemOpen
  \bibfield  {author} {\bibinfo {author} {\bibfnamefont {L.}~\bibnamefont
  {Rio}}, \bibinfo {author} {\bibfnamefont {J.}~\bibnamefont {Åberg}},
  \bibinfo {author} {\bibfnamefont {R.}~\bibnamefont {Renner}}, \bibinfo
  {author} {\bibfnamefont {O.}~\bibnamefont {Dahlsten}}, \ and\ \bibinfo
  {author} {\bibfnamefont {V.}~\bibnamefont {Vedral}},\ }\bibfield  {title}
  {\enquote {\bibinfo {title} {The thermodynamic meaning of negative
  entropy},}\ }\href {https://doi.org/10.1038/nature10123} {\bibfield
  {journal} {\bibinfo  {journal} {Nature}\ }\textbf {\bibinfo {volume} {474}},\
  \bibinfo {pages} {61} (\bibinfo {year} {2011})}\BibitemShut {NoStop}%
\bibitem [{\citenamefont {Groisman}\ \emph {et~al.}(2005)\citenamefont
  {Groisman}, \citenamefont {Popescu},\ and\ \citenamefont
  {Winter}}]{Groisman.05.PRA}%
  \BibitemOpen
  \bibfield  {author} {\bibinfo {author} {\bibfnamefont {B.}~\bibnamefont
  {Groisman}}, \bibinfo {author} {\bibfnamefont {S.}~\bibnamefont {Popescu}}, \
  and\ \bibinfo {author} {\bibfnamefont {A.}~\bibnamefont {Winter}},\
  }\bibfield  {title} {\enquote {\bibinfo {title} {Quantum, classical, and
  total amount of correlations in a quantum state},}\ }\href {\doibase
  10.1103/PhysRevA.72.032317} {\bibfield  {journal} {\bibinfo  {journal} {Phys.
  Rev. A}\ }\textbf {\bibinfo {volume} {72}},\ \bibinfo {pages} {032317}
  (\bibinfo {year} {2005})}\BibitemShut {NoStop}%
\bibitem [{\citenamefont {Modi}\ \emph {et~al.}(2012)\citenamefont {Modi},
  \citenamefont {Brodutch}, \citenamefont {Cable}, \citenamefont {Paterek},\
  and\ \citenamefont {Vedral}}]{Modi.12.RMP}%
  \BibitemOpen
  \bibfield  {author} {\bibinfo {author} {\bibfnamefont {K.}~\bibnamefont
  {Modi}}, \bibinfo {author} {\bibfnamefont {A.}~\bibnamefont {Brodutch}},
  \bibinfo {author} {\bibfnamefont {H.}~\bibnamefont {Cable}}, \bibinfo
  {author} {\bibfnamefont {T.}~\bibnamefont {Paterek}}, \ and\ \bibinfo
  {author} {\bibfnamefont {V.}~\bibnamefont {Vedral}},\ }\bibfield  {title}
  {\enquote {\bibinfo {title} {The classical-quantum boundary for correlations:
  Discord and related measures},}\ }\href {\doibase 10.1103/RevModPhys.84.1655}
  {\bibfield  {journal} {\bibinfo  {journal} {Rev. Mod. Phys.}\ }\textbf
  {\bibinfo {volume} {84}},\ \bibinfo {pages} {1655} (\bibinfo {year}
  {2012})}\BibitemShut {NoStop}%
\bibitem [{\citenamefont {Crooks}(1999)}]{Crooks.99.PRE}%
  \BibitemOpen
  \bibfield  {author} {\bibinfo {author} {\bibfnamefont {G.}~\bibnamefont
  {Crooks}},\ }\bibfield  {title} {\enquote {\bibinfo {title} {Entropy
  production fluctuation theorem and the nonequilibrium work relation for free
  energy differences},}\ }\href {\doibase 10.1103/PhysRevE.60.2721} {\bibfield
  {journal} {\bibinfo  {journal} {Phys. Rev. E}\ }\textbf {\bibinfo {volume}
  {60}},\ \bibinfo {pages} {2721} (\bibinfo {year} {1999})}\BibitemShut
  {NoStop}%
\bibitem [{\citenamefont {Vaikuntanathan}\ and\ \citenamefont
  {Jarzynski}(2009)}]{Vaikuntanathan.09.EL}%
  \BibitemOpen
  \bibfield  {author} {\bibinfo {author} {\bibfnamefont {S.}~\bibnamefont
  {Vaikuntanathan}}\ and\ \bibinfo {author} {\bibfnamefont {C.}~\bibnamefont
  {Jarzynski}},\ }\bibfield  {title} {\enquote {\bibinfo {title} {Dissipation
  and lag in irreversible processes},}\ }\href
  {http://stacks.iop.org/0295-5075/87/i=6/a=60005} {\bibfield  {journal}
  {\bibinfo  {journal} {Europhys. Lett.}\ }\textbf {\bibinfo {volume} {87}},\
  \bibinfo {pages} {60005} (\bibinfo {year} {2009})}\BibitemShut {NoStop}%
\bibitem [{\citenamefont {Sagawa}\ and\ \citenamefont
  {Ueda}(2013)}]{Sagawa.13.NJP}%
  \BibitemOpen
  \bibfield  {author} {\bibinfo {author} {\bibfnamefont {T.}~\bibnamefont
  {Sagawa}}\ and\ \bibinfo {author} {\bibfnamefont {M.}~\bibnamefont {Ueda}},\
  }\bibfield  {title} {\enquote {\bibinfo {title} {Role of mutual information
  in entropy production under information exchanges},}\ }\href {\doibase
  10.1088/1367-2630/15/12/125012} {\bibfield  {journal} {\bibinfo  {journal}
  {New Journal of Physics}\ }\textbf {\bibinfo {volume} {15}},\ \bibinfo
  {pages} {125012} (\bibinfo {year} {2013})}\BibitemShut {NoStop}%
\bibitem [{\citenamefont {Barato}\ \emph {et~al.}(2013)\citenamefont {Barato},
  \citenamefont {Hartich},\ and\ \citenamefont {Seifert}}]{Barato.13.PRE}%
  \BibitemOpen
  \bibfield  {author} {\bibinfo {author} {\bibfnamefont {A.}~\bibnamefont
  {Barato}}, \bibinfo {author} {\bibfnamefont {D.}~\bibnamefont {Hartich}}, \
  and\ \bibinfo {author} {\bibfnamefont {U.}~\bibnamefont {Seifert}},\
  }\bibfield  {title} {\enquote {\bibinfo {title} {Information-theoretic versus
  thermodynamic entropy production in autonomous sensory networks},}\ }\href
  {\doibase 10.1103/PhysRevE.87.042104} {\bibfield  {journal} {\bibinfo
  {journal} {Phys. Rev. E}\ }\textbf {\bibinfo {volume} {87}},\ \bibinfo
  {pages} {042104} (\bibinfo {year} {2013})}\BibitemShut {NoStop}%
\bibitem [{\citenamefont {Horowitz}\ and\ \citenamefont
  {Esposito}(2014)}]{Horowitz.14.PRX}%
  \BibitemOpen
  \bibfield  {author} {\bibinfo {author} {\bibfnamefont {J.}~\bibnamefont
  {Horowitz}}\ and\ \bibinfo {author} {\bibfnamefont {M.}~\bibnamefont
  {Esposito}},\ }\bibfield  {title} {\enquote {\bibinfo {title} {Thermodynamics
  with continuous information flow},}\ }\href {\doibase
  10.1103/PhysRevX.4.031015} {\bibfield  {journal} {\bibinfo  {journal} {Phys.
  Rev. X}\ }\textbf {\bibinfo {volume} {4}},\ \bibinfo {pages} {031015}
  (\bibinfo {year} {2014})}\BibitemShut {NoStop}%
\bibitem [{\citenamefont {Bruch}\ \emph {et~al.}(2016)\citenamefont {Bruch},
  \citenamefont {Thomas}, \citenamefont {Viola~Kusminskiy}, \citenamefont {von
  Oppen},\ and\ \citenamefont {Nitzan}}]{Bruch.16.PRB}%
  \BibitemOpen
  \bibfield  {author} {\bibinfo {author} {\bibfnamefont {A.}~\bibnamefont
  {Bruch}}, \bibinfo {author} {\bibfnamefont {M.}~\bibnamefont {Thomas}},
  \bibinfo {author} {\bibfnamefont {S.}~\bibnamefont {Viola~Kusminskiy}},
  \bibinfo {author} {\bibfnamefont {F.}~\bibnamefont {von Oppen}}, \ and\
  \bibinfo {author} {\bibfnamefont {A.}~\bibnamefont {Nitzan}},\ }\bibfield
  {title} {\enquote {\bibinfo {title} {Quantum thermodynamics of the driven
  resonant level model},}\ }\href {\doibase 10.1103/PhysRevB.93.115318}
  {\bibfield  {journal} {\bibinfo  {journal} {Phys. Rev. B}\ }\textbf {\bibinfo
  {volume} {93}},\ \bibinfo {pages} {115318} (\bibinfo {year}
  {2016})}\BibitemShut {NoStop}%
\bibitem [{\citenamefont {Cangemi}\ \emph {et~al.}(2021)\citenamefont
  {Cangemi}, \citenamefont {Carrega}, \citenamefont {De~Candia}, \citenamefont
  {Cataudella}, \citenamefont {De~Filippis}, \citenamefont {Sassetti},\ and\
  \citenamefont {Benenti}}]{Cangemi.21.PRR}%
  \BibitemOpen
  \bibfield  {author} {\bibinfo {author} {\bibfnamefont {L.}~\bibnamefont
  {Cangemi}}, \bibinfo {author} {\bibfnamefont {M.}~\bibnamefont {Carrega}},
  \bibinfo {author} {\bibfnamefont {A.}~\bibnamefont {De~Candia}}, \bibinfo
  {author} {\bibfnamefont {V.}~\bibnamefont {Cataudella}}, \bibinfo {author}
  {\bibfnamefont {G.}~\bibnamefont {De~Filippis}}, \bibinfo {author}
  {\bibfnamefont {M.}~\bibnamefont {Sassetti}}, \ and\ \bibinfo {author}
  {\bibfnamefont {G.}~\bibnamefont {Benenti}},\ }\bibfield  {title} {\enquote
  {\bibinfo {title} {Optimal energy conversion through antiadiabatic driving
  breaking time-reversal symmetry},}\ }\href {\doibase
  10.1103/PhysRevResearch.3.013237} {\bibfield  {journal} {\bibinfo  {journal}
  {Phys. Rev. Research}\ }\textbf {\bibinfo {volume} {3}},\ \bibinfo {pages}
  {013237} (\bibinfo {year} {2021})}\BibitemShut {NoStop}%
\bibitem [{\citenamefont {Liu}\ \emph {et~al.}(2021)\citenamefont {Liu},
  \citenamefont {Jung},\ and\ \citenamefont {Segal}}]{Liu.21.PRL}%
  \BibitemOpen
  \bibfield  {author} {\bibinfo {author} {\bibfnamefont {J.}~\bibnamefont
  {Liu}}, \bibinfo {author} {\bibfnamefont {K.}~\bibnamefont {Jung}}, \ and\
  \bibinfo {author} {\bibfnamefont {D.}~\bibnamefont {Segal}},\ }\bibfield
  {title} {\enquote {\bibinfo {title} {Periodically driven quantum thermal
  machines from warming up to limit cycle},}\ }\href {\doibase
  10.1103/PhysRevLett.127.200602} {\bibfield  {journal} {\bibinfo  {journal}
  {Phys. Rev. Lett.}\ }\textbf {\bibinfo {volume} {127}},\ \bibinfo {pages}
  {200602} (\bibinfo {year} {2021})}\BibitemShut {NoStop}%
\bibitem [{\citenamefont {Deffner}\ and\ \citenamefont
  {Jarzynski}(2013)}]{Deffner.13.PRX}%
  \BibitemOpen
  \bibfield  {author} {\bibinfo {author} {\bibfnamefont {S.}~\bibnamefont
  {Deffner}}\ and\ \bibinfo {author} {\bibfnamefont {C.}~\bibnamefont
  {Jarzynski}},\ }\bibfield  {title} {\enquote {\bibinfo {title} {Information
  processing and the second law of thermodynamics: An inclusive, hamiltonian
  approach},}\ }\href {\doibase 10.1103/PhysRevX.3.041003} {\bibfield
  {journal} {\bibinfo  {journal} {Phys. Rev. X}\ }\textbf {\bibinfo {volume}
  {3}},\ \bibinfo {pages} {041003} (\bibinfo {year} {2013})}\BibitemShut
  {NoStop}%
\bibitem [{\citenamefont {Liu}\ and\ \citenamefont {Nie}(2023)}]{Liu.23.PRAa}%
  \BibitemOpen
  \bibfield  {author} {\bibinfo {author} {\bibfnamefont {J.}~\bibnamefont
  {Liu}}\ and\ \bibinfo {author} {\bibfnamefont {H.}~\bibnamefont {Nie}},\
  }\bibfield  {title} {\enquote {\bibinfo {title} {Universal landauer-like
  inequality from the first law of thermodynamics},}\ }\href {\doibase
  10.1103/PhysRevA.108.L040203} {\bibfield  {journal} {\bibinfo  {journal}
  {Phys. Rev. A}\ }\textbf {\bibinfo {volume} {108}},\ \bibinfo {pages}
  {L040203} (\bibinfo {year} {2023})}\BibitemShut {NoStop}%
\bibitem [{SM()}]{SM}%
  \BibitemOpen
  \href@noop {} {}\bibinfo {note} {See Supplemental Material for additional
  theoretical analysis as well as numerical results that complement the main
  text.}\BibitemShut {Stop}%
\bibitem [{\citenamefont {Miller}\ \emph {et~al.}(2020)\citenamefont {Miller},
  \citenamefont {Guarnieri}, \citenamefont {Mitchison},\ and\ \citenamefont
  {Goold}}]{Miller.20.PRL}%
  \BibitemOpen
  \bibfield  {author} {\bibinfo {author} {\bibfnamefont {H.}~\bibnamefont
  {Miller}}, \bibinfo {author} {\bibfnamefont {G.}~\bibnamefont {Guarnieri}},
  \bibinfo {author} {\bibfnamefont {M.}~\bibnamefont {Mitchison}}, \ and\
  \bibinfo {author} {\bibfnamefont {J.}~\bibnamefont {Goold}},\ }\bibfield
  {title} {\enquote {\bibinfo {title} {Quantum fluctuations hinder finite-time
  information erasure near the landauer limit},}\ }\href {\doibase
  10.1103/PhysRevLett.125.160602} {\bibfield  {journal} {\bibinfo  {journal}
  {Phys. Rev. Lett.}\ }\textbf {\bibinfo {volume} {125}},\ \bibinfo {pages}
  {160602} (\bibinfo {year} {2020})}\BibitemShut {NoStop}%
\bibitem [{\citenamefont {Riechers}\ and\ \citenamefont
  {Gu}(2021)}]{Riechers.21.PRA}%
  \BibitemOpen
  \bibfield  {author} {\bibinfo {author} {\bibfnamefont {P.}~\bibnamefont
  {Riechers}}\ and\ \bibinfo {author} {\bibfnamefont {M.}~\bibnamefont {Gu}},\
  }\bibfield  {title} {\enquote {\bibinfo {title} {Impossibility of achieving
  landauer's bound for almost every quantum state},}\ }\href {\doibase
  10.1103/PhysRevA.104.012214} {\bibfield  {journal} {\bibinfo  {journal}
  {Phys. Rev. A}\ }\textbf {\bibinfo {volume} {104}},\ \bibinfo {pages}
  {012214} (\bibinfo {year} {2021})}\BibitemShut {NoStop}%
\bibitem [{\citenamefont {Kandel}\ \emph {et~al.}(2019)\citenamefont {Kandel},
  \citenamefont {Qiao}, \citenamefont {Fallahi}, \citenamefont {Gardner},
  \citenamefont {Manfra},\ and\ \citenamefont {Nichol}}]{Kandel.19.N}%
  \BibitemOpen
  \bibfield  {author} {\bibinfo {author} {\bibfnamefont {Y.}~\bibnamefont
  {Kandel}}, \bibinfo {author} {\bibfnamefont {H.}~\bibnamefont {Qiao}},
  \bibinfo {author} {\bibfnamefont {S.}~\bibnamefont {Fallahi}}, \bibinfo
  {author} {\bibfnamefont {G.}~\bibnamefont {Gardner}}, \bibinfo {author}
  {\bibfnamefont {M.}~\bibnamefont {Manfra}}, \ and\ \bibinfo {author}
  {\bibfnamefont {J.}~\bibnamefont {Nichol}},\ }\bibfield  {title} {\enquote
  {\bibinfo {title} {Coherent spin-state transfer via heisenberg exchange},}\
  }\href {https://doi.org/10.1038/s41586-019-1566-8} {\bibfield  {journal}
  {\bibinfo  {journal} {Nature}\ }\textbf {\bibinfo {volume} {573}},\ \bibinfo
  {pages} {553} (\bibinfo {year} {2019})}\BibitemShut {NoStop}%
\bibitem [{\citenamefont {Karamlou}\ \emph {et~al.}(2022)\citenamefont
  {Karamlou}, \citenamefont {Braum\"uller}, \citenamefont {Yanay},
  \citenamefont {Di~Paolo}, \citenamefont {Harrington}, \citenamefont {Kannan},
  \citenamefont {Kim}, \citenamefont {Kjaergaard}, \citenamefont {Melville},
  \citenamefont {Muschinske}, \citenamefont {Niedzielski}, \citenamefont
  {Veps\"al\"ainen}, \citenamefont {Winik}, \citenamefont {Yoder},
  \citenamefont {Schwartz}, \citenamefont {Tahan}, \citenamefont {Orlando},
  \citenamefont {Gustavsson},\ and\ \citenamefont {Oliver}}]{Karamlou.22.NPJQ}%
  \BibitemOpen
  \bibfield  {author} {\bibinfo {author} {\bibfnamefont {A.}~\bibnamefont
  {Karamlou}}, \bibinfo {author} {\bibfnamefont {J.}~\bibnamefont
  {Braum\"uller}}, \bibinfo {author} {\bibfnamefont {Y.}~\bibnamefont {Yanay}},
  \bibinfo {author} {\bibfnamefont {A.}~\bibnamefont {Di~Paolo}}, \bibinfo
  {author} {\bibfnamefont {P.}~\bibnamefont {Harrington}}, \bibinfo {author}
  {\bibfnamefont {B.}~\bibnamefont {Kannan}}, \bibinfo {author} {\bibfnamefont
  {D.}~\bibnamefont {Kim}}, \bibinfo {author} {\bibfnamefont {M.}~\bibnamefont
  {Kjaergaard}}, \bibinfo {author} {\bibfnamefont {A.}~\bibnamefont
  {Melville}}, \bibinfo {author} {\bibfnamefont {S.}~\bibnamefont
  {Muschinske}}, \bibinfo {author} {\bibfnamefont {B.}~\bibnamefont
  {Niedzielski}}, \bibinfo {author} {\bibfnamefont {A.}~\bibnamefont
  {Veps\"al\"ainen}}, \bibinfo {author} {\bibfnamefont {R.}~\bibnamefont
  {Winik}}, \bibinfo {author} {\bibfnamefont {J.}~\bibnamefont {Yoder}},
  \bibinfo {author} {\bibfnamefont {M.}~\bibnamefont {Schwartz}}, \bibinfo
  {author} {\bibfnamefont {C.}~\bibnamefont {Tahan}}, \bibinfo {author}
  {\bibfnamefont {T.}~\bibnamefont {Orlando}}, \bibinfo {author} {\bibfnamefont
  {S.}~\bibnamefont {Gustavsson}}, \ and\ \bibinfo {author} {\bibfnamefont
  {W.}~\bibnamefont {Oliver}},\ }\bibfield  {title} {\enquote {\bibinfo {title}
  {Quantum transport and localization in 1d and 2d tight-binding lattices},}\
  }\href {https://doi.org/10.1038/s41534-022-00528-0} {\bibfield  {journal}
  {\bibinfo  {journal} {npj Quantum Inf.}\ }\textbf {\bibinfo {volume} {8}},\
  \bibinfo {pages} {35} (\bibinfo {year} {2022})}\BibitemShut {NoStop}%
\bibitem [{\citenamefont {DiVincenzo}(2000)}]{DiVincenzo.00.FP}%
  \BibitemOpen
  \bibfield  {author} {\bibinfo {author} {\bibfnamefont {D.}~\bibnamefont
  {DiVincenzo}},\ }\bibfield  {title} {\enquote {\bibinfo {title} {The physical
  implementation of quantum computation},}\ }\href {\doibase
  https://doi.org/10.1002/1521-3978(200009)48:9/11<771::AID-PROP771>3.0.CO;2-E}
  {\bibfield  {journal} {\bibinfo  {journal} {Fortschr. Phys.}\ }\textbf
  {\bibinfo {volume} {48}},\ \bibinfo {pages} {771} (\bibinfo {year}
  {2000})}\BibitemShut {NoStop}%
\bibitem [{\citenamefont {Schindler}\ \emph {et~al.}(2011)\citenamefont
  {Schindler}, \citenamefont {Barreiro}, \citenamefont {Monz}, \citenamefont
  {Nebendahl}, \citenamefont {Nigg}, \citenamefont {Chwalla}, \citenamefont
  {Hennrich},\ and\ \citenamefont {Blatt}}]{Schindler.11.S}%
  \BibitemOpen
  \bibfield  {author} {\bibinfo {author} {\bibfnamefont {P.}~\bibnamefont
  {Schindler}}, \bibinfo {author} {\bibfnamefont {J.}~\bibnamefont {Barreiro}},
  \bibinfo {author} {\bibfnamefont {T.}~\bibnamefont {Monz}}, \bibinfo {author}
  {\bibfnamefont {V.}~\bibnamefont {Nebendahl}}, \bibinfo {author}
  {\bibfnamefont {D.}~\bibnamefont {Nigg}}, \bibinfo {author} {\bibfnamefont
  {M.}~\bibnamefont {Chwalla}}, \bibinfo {author} {\bibfnamefont
  {M.}~\bibnamefont {Hennrich}}, \ and\ \bibinfo {author} {\bibfnamefont
  {R.}~\bibnamefont {Blatt}},\ }\bibfield  {title} {\enquote {\bibinfo {title}
  {Experimental repetitive quantum error correction},}\ }\href {\doibase
  10.1126/science.1203329} {\bibfield  {journal} {\bibinfo  {journal}
  {Science}\ }\textbf {\bibinfo {volume} {332}},\ \bibinfo {pages} {1059}
  (\bibinfo {year} {2011})}\BibitemShut {NoStop}%
\bibitem [{\citenamefont {Reed}\ \emph {et~al.}(2012)\citenamefont {Reed},
  \citenamefont {DiCarlo}, \citenamefont {Nigg}, \citenamefont {Sun},
  \citenamefont {Frunzio}, \citenamefont {Girvin},\ and\ \citenamefont
  {Schoelkopf}}]{Reed.12.N}%
  \BibitemOpen
  \bibfield  {author} {\bibinfo {author} {\bibfnamefont {M.}~\bibnamefont
  {Reed}}, \bibinfo {author} {\bibfnamefont {L.}~\bibnamefont {DiCarlo}},
  \bibinfo {author} {\bibfnamefont {S.~E.}\ \bibnamefont {Nigg}}, \bibinfo
  {author} {\bibfnamefont {L.}~\bibnamefont {Sun}}, \bibinfo {author}
  {\bibfnamefont {L.}~\bibnamefont {Frunzio}}, \bibinfo {author} {\bibfnamefont
  {S.}~\bibnamefont {Girvin}}, \ and\ \bibinfo {author} {\bibfnamefont
  {R.}~\bibnamefont {Schoelkopf}},\ }\bibfield  {title} {\enquote {\bibinfo
  {title} {Realization of three-qubit quantum error correction with
  superconducting circuits},}\ }\href {\doibase 10.1038/nature10786} {\bibfield
   {journal} {\bibinfo  {journal} {Nature}\ }\textbf {\bibinfo {volume}
  {482}},\ \bibinfo {pages} {382} (\bibinfo {year} {2012})}\BibitemShut
  {NoStop}%
\bibitem [{\citenamefont {Cattaneo}\ \emph {et~al.}(2019)\citenamefont
  {Cattaneo}, \citenamefont {Giorgi}, \citenamefont {Maniscalco},\ and\
  \citenamefont {Zambrini}}]{Cattaneo.19.NJP}%
  \BibitemOpen
  \bibfield  {author} {\bibinfo {author} {\bibfnamefont {M.}~\bibnamefont
  {Cattaneo}}, \bibinfo {author} {\bibfnamefont {G.}~\bibnamefont {Giorgi}},
  \bibinfo {author} {\bibfnamefont {S.}~\bibnamefont {Maniscalco}}, \ and\
  \bibinfo {author} {\bibfnamefont {R.}~\bibnamefont {Zambrini}},\ }\bibfield
  {title} {\enquote {\bibinfo {title} {Local versus global master equation with
  common and separate baths: superiority of the global approach in partial
  secular approximation},}\ }\href {\doibase 10.1088/1367-2630/ab54ac}
  {\bibfield  {journal} {\bibinfo  {journal} {New J. Phys.}\ }\textbf {\bibinfo
  {volume} {21}},\ \bibinfo {pages} {113045} (\bibinfo {year}
  {2019})}\BibitemShut {NoStop}%
\bibitem [{\citenamefont {Lanyon}\ \emph {et~al.}(2017)\citenamefont {Lanyon},
  \citenamefont {Maier}, \citenamefont {Holz{\"a}pfel}, \citenamefont
  {Baumgratz}, \citenamefont {Hempel}, \citenamefont {Jurcevic}, \citenamefont
  {Dhand}, \citenamefont {Buyskikh}, \citenamefont {Daley}, \citenamefont
  {Cramer}, \citenamefont {Plenio}, \citenamefont {Blatt},\ and\ \citenamefont
  {Roos}}]{Lanyon.17.NP}%
  \BibitemOpen
  \bibfield  {author} {\bibinfo {author} {\bibfnamefont {B.}~\bibnamefont
  {Lanyon}}, \bibinfo {author} {\bibfnamefont {C.}~\bibnamefont {Maier}},
  \bibinfo {author} {\bibfnamefont {M.}~\bibnamefont {Holz{\"a}pfel}}, \bibinfo
  {author} {\bibfnamefont {T.}~\bibnamefont {Baumgratz}}, \bibinfo {author}
  {\bibfnamefont {C.}~\bibnamefont {Hempel}}, \bibinfo {author} {\bibfnamefont
  {P.}~\bibnamefont {Jurcevic}}, \bibinfo {author} {\bibfnamefont
  {I.}~\bibnamefont {Dhand}}, \bibinfo {author} {\bibfnamefont {A.~S.}\
  \bibnamefont {Buyskikh}}, \bibinfo {author} {\bibfnamefont {A.~J.}\
  \bibnamefont {Daley}}, \bibinfo {author} {\bibfnamefont {M.}~\bibnamefont
  {Cramer}}, \bibinfo {author} {\bibfnamefont {M.}~\bibnamefont {Plenio}},
  \bibinfo {author} {\bibfnamefont {R.}~\bibnamefont {Blatt}}, \ and\ \bibinfo
  {author} {\bibfnamefont {C.}~\bibnamefont {Roos}},\ }\bibfield  {title}
  {\enquote {\bibinfo {title} {Efficient tomography of a quantum many-body
  system},}\ }\href {\doibase 10.1038/nphys4244} {\bibfield  {journal}
  {\bibinfo  {journal} {Nat. Phys.}\ }\textbf {\bibinfo {volume} {13}},\
  \bibinfo {pages} {1158} (\bibinfo {year} {2017})}\BibitemShut {NoStop}%
\bibitem [{\citenamefont {Yoneda}\ \emph {et~al.}(2021)\citenamefont {Yoneda},
  \citenamefont {Huang}, \citenamefont {Feng}, \citenamefont {Yang},
  \citenamefont {Chan}, \citenamefont {Tanttu}, \citenamefont {Gilbert},
  \citenamefont {Leon}, \citenamefont {Hudson}, \citenamefont {Itoh},
  \citenamefont {Morello}, \citenamefont {Bartlett}, \citenamefont {Laucht},
  \citenamefont {Saraiva},\ and\ \citenamefont {Dzurak}}]{Yoneda.21.NC}%
  \BibitemOpen
  \bibfield  {author} {\bibinfo {author} {\bibfnamefont {J.}~\bibnamefont
  {Yoneda}}, \bibinfo {author} {\bibfnamefont {W.}~\bibnamefont {Huang}},
  \bibinfo {author} {\bibfnamefont {M.}~\bibnamefont {Feng}}, \bibinfo {author}
  {\bibfnamefont {C.}~\bibnamefont {Yang}}, \bibinfo {author} {\bibfnamefont
  {K.}~\bibnamefont {Chan}}, \bibinfo {author} {\bibfnamefont {T.}~\bibnamefont
  {Tanttu}}, \bibinfo {author} {\bibfnamefont {W.}~\bibnamefont {Gilbert}},
  \bibinfo {author} {\bibfnamefont {R.}~\bibnamefont {Leon}}, \bibinfo {author}
  {\bibfnamefont {F.}~\bibnamefont {Hudson}}, \bibinfo {author} {\bibfnamefont
  {K.}~\bibnamefont {Itoh}}, \bibinfo {author} {\bibfnamefont {A.}~\bibnamefont
  {Morello}}, \bibinfo {author} {\bibfnamefont {S.}~\bibnamefont {Bartlett}},
  \bibinfo {author} {\bibfnamefont {A.}~\bibnamefont {Laucht}}, \bibinfo
  {author} {\bibfnamefont {A.}~\bibnamefont {Saraiva}}, \ and\ \bibinfo
  {author} {\bibfnamefont {A.}~\bibnamefont {Dzurak}},\ }\bibfield  {title}
  {\enquote {\bibinfo {title} {Coherent spin qubit transport in silicon},}\
  }\href {https://doi.org/10.1038/s41467-021-24371-7} {\bibfield  {journal}
  {\bibinfo  {journal} {Nat. Commun.}\ }\textbf {\bibinfo {volume} {12}},\
  \bibinfo {pages} {4114} (\bibinfo {year} {2021})}\BibitemShut {NoStop}%
\bibitem [{\citenamefont {Zhong}\ \emph {et~al.}(2021)\citenamefont {Zhong},
  \citenamefont {Chang}, \citenamefont {Bienfait}, \citenamefont {Dumur},
  \citenamefont {Chou}, \citenamefont {Conner}, \citenamefont {Grebel},
  \citenamefont {Povey}, \citenamefont {Yan}, \citenamefont {Schuster},\ and\
  \citenamefont {Cleland}}]{Zhong.21.N}%
  \BibitemOpen
  \bibfield  {author} {\bibinfo {author} {\bibfnamefont {Y.}~\bibnamefont
  {Zhong}}, \bibinfo {author} {\bibfnamefont {H.}~\bibnamefont {Chang}},
  \bibinfo {author} {\bibfnamefont {A.}~\bibnamefont {Bienfait}}, \bibinfo
  {author} {\bibfnamefont {E.}~\bibnamefont {Dumur}}, \bibinfo {author}
  {\bibfnamefont {M.}~\bibnamefont {Chou}}, \bibinfo {author} {\bibfnamefont
  {C.}~\bibnamefont {Conner}}, \bibinfo {author} {\bibfnamefont
  {J.}~\bibnamefont {Grebel}}, \bibinfo {author} {\bibfnamefont
  {R.}~\bibnamefont {Povey}}, \bibinfo {author} {\bibfnamefont
  {H.}~\bibnamefont {Yan}}, \bibinfo {author} {\bibfnamefont {D.}~\bibnamefont
  {Schuster}}, \ and\ \bibinfo {author} {\bibfnamefont {A.}~\bibnamefont
  {Cleland}},\ }\bibfield  {title} {\enquote {\bibinfo {title} {Deterministic
  multi-qubit entanglement in a quantum network},}\ }\href
  {https://doi.org/10.1038/s41586-021-03288-7} {\bibfield  {journal} {\bibinfo
  {journal} {Nature}\ }\textbf {\bibinfo {volume} {590}},\ \bibinfo {pages}
  {571} (\bibinfo {year} {2021})}\BibitemShut {NoStop}%
\bibitem [{\citenamefont {Torlai}\ \emph {et~al.}(2018)\citenamefont {Torlai},
  \citenamefont {Mazzola}, \citenamefont {Carrasquilla}, \citenamefont
  {Troyer}, \citenamefont {Melko},\ and\ \citenamefont
  {Carleo}}]{Torlai.18.NP}%
  \BibitemOpen
  \bibfield  {author} {\bibinfo {author} {\bibfnamefont {G.}~\bibnamefont
  {Torlai}}, \bibinfo {author} {\bibfnamefont {G.}~\bibnamefont {Mazzola}},
  \bibinfo {author} {\bibfnamefont {J.}~\bibnamefont {Carrasquilla}}, \bibinfo
  {author} {\bibfnamefont {M.}~\bibnamefont {Troyer}}, \bibinfo {author}
  {\bibfnamefont {R.}~\bibnamefont {Melko}}, \ and\ \bibinfo {author}
  {\bibfnamefont {G.}~\bibnamefont {Carleo}},\ }\bibfield  {title} {\enquote
  {\bibinfo {title} {Neural-network quantum state tomography},}\ }\href
  {\doibase 10.1038/s41567-018-0048-5} {\bibfield  {journal} {\bibinfo
  {journal} {Nat. Phys.}\ }\textbf {\bibinfo {volume} {14}},\ \bibinfo {pages}
  {447} (\bibinfo {year} {2018})}\BibitemShut {NoStop}%
\bibitem [{\citenamefont {Ahmed}\ \emph {et~al.}(2021)\citenamefont {Ahmed},
  \citenamefont {S\'anchez Mu\~noz}, \citenamefont {Nori},\ and\ \citenamefont
  {Kockum}}]{Ahmed.21.PRL}%
  \BibitemOpen
  \bibfield  {author} {\bibinfo {author} {\bibfnamefont {S.}~\bibnamefont
  {Ahmed}}, \bibinfo {author} {\bibfnamefont {C.}~\bibnamefont {S\'anchez
  Mu\~noz}}, \bibinfo {author} {\bibfnamefont {F.}~\bibnamefont {Nori}}, \ and\
  \bibinfo {author} {\bibfnamefont {A.}~\bibnamefont {Kockum}},\ }\bibfield
  {title} {\enquote {\bibinfo {title} {Quantum state tomography with
  conditional generative adversarial networks},}\ }\href {\doibase
  10.1103/PhysRevLett.127.140502} {\bibfield  {journal} {\bibinfo  {journal}
  {Phys. Rev. Lett.}\ }\textbf {\bibinfo {volume} {127}},\ \bibinfo {pages}
  {140502} (\bibinfo {year} {2021})}\BibitemShut {NoStop}%
\bibitem [{\citenamefont {Struchalin}\ \emph {et~al.}(2021)\citenamefont
  {Struchalin}, \citenamefont {Zagorovskii}, \citenamefont {Kovlakov},
  \citenamefont {Straupe},\ and\ \citenamefont {Kulik}}]{Struchalin.21.PRXQ}%
  \BibitemOpen
  \bibfield  {author} {\bibinfo {author} {\bibfnamefont {G.}~\bibnamefont
  {Struchalin}}, \bibinfo {author} {\bibfnamefont {Ya.}\ \bibnamefont
  {Zagorovskii}}, \bibinfo {author} {\bibfnamefont {E.}~\bibnamefont
  {Kovlakov}}, \bibinfo {author} {\bibfnamefont {S.}~\bibnamefont {Straupe}}, \
  and\ \bibinfo {author} {\bibfnamefont {S.}~\bibnamefont {Kulik}},\ }\bibfield
   {title} {\enquote {\bibinfo {title} {Experimental estimation of quantum
  state properties from classical shadows},}\ }\href {\doibase
  10.1103/PRXQuantum.2.010307} {\bibfield  {journal} {\bibinfo  {journal} {PRX
  Quantum}\ }\textbf {\bibinfo {volume} {2}},\ \bibinfo {pages} {010307}
  (\bibinfo {year} {2021})}\BibitemShut {NoStop}%
\bibitem [{\citenamefont {Schaller}\ \emph {et~al.}(2014)\citenamefont
  {Schaller}, \citenamefont {Nietner},\ and\ \citenamefont
  {Brandes}}]{Schaller.14.NJP}%
  \BibitemOpen
  \bibfield  {author} {\bibinfo {author} {\bibfnamefont {G.}~\bibnamefont
  {Schaller}}, \bibinfo {author} {\bibfnamefont {C.}~\bibnamefont {Nietner}}, \
  and\ \bibinfo {author} {\bibfnamefont {T.}~\bibnamefont {Brandes}},\
  }\bibfield  {title} {\enquote {\bibinfo {title} {Relaxation dynamics of
  meso-reservoirs},}\ }\href {\doibase 10.1088/1367-2630/16/12/125011}
  {\bibfield  {journal} {\bibinfo  {journal} {New J. Phys.}\ }\textbf {\bibinfo
  {volume} {16}},\ \bibinfo {pages} {125011} (\bibinfo {year}
  {2014})}\BibitemShut {NoStop}%
\bibitem [{\citenamefont {Amato}\ \emph {et~al.}(2020)\citenamefont {Amato},
  \citenamefont {Breuer}, \citenamefont {Wimberger}, \citenamefont
  {Rodr\'{\i}guez},\ and\ \citenamefont {Buchleitner}}]{Amato.20.PRA}%
  \BibitemOpen
  \bibfield  {author} {\bibinfo {author} {\bibfnamefont {G.}~\bibnamefont
  {Amato}}, \bibinfo {author} {\bibfnamefont {H.-P.}\ \bibnamefont {Breuer}},
  \bibinfo {author} {\bibfnamefont {S.}~\bibnamefont {Wimberger}}, \bibinfo
  {author} {\bibfnamefont {A.}~\bibnamefont {Rodr\'{\i}guez}}, \ and\ \bibinfo
  {author} {\bibfnamefont {A.}~\bibnamefont {Buchleitner}},\ }\bibfield
  {title} {\enquote {\bibinfo {title} {Noninteracting many-particle quantum
  transport between finite reservoirs},}\ }\href {\doibase
  10.1103/PhysRevA.102.022207} {\bibfield  {journal} {\bibinfo  {journal}
  {Phys. Rev. A}\ }\textbf {\bibinfo {volume} {102}},\ \bibinfo {pages}
  {022207} (\bibinfo {year} {2020})}\BibitemShut {NoStop}%
\bibitem [{\citenamefont {Riera-Campeny}\ \emph {et~al.}(2021)\citenamefont
  {Riera-Campeny}, \citenamefont {Sanpera},\ and\ \citenamefont
  {Strasberg}}]{Campeny.21.PRXQ}%
  \BibitemOpen
  \bibfield  {author} {\bibinfo {author} {\bibfnamefont {A.}~\bibnamefont
  {Riera-Campeny}}, \bibinfo {author} {\bibfnamefont {A.}~\bibnamefont
  {Sanpera}}, \ and\ \bibinfo {author} {\bibfnamefont {P.}~\bibnamefont
  {Strasberg}},\ }\bibfield  {title} {\enquote {\bibinfo {title} {Quantum
  systems correlated with a finite bath: Nonequilibrium dynamics and
  thermodynamics},}\ }\href {\doibase 10.1103/PRXQuantum.2.010340} {\bibfield
  {journal} {\bibinfo  {journal} {PRX Quantum}\ }\textbf {\bibinfo {volume}
  {2}},\ \bibinfo {pages} {010340} (\bibinfo {year} {2021})}\BibitemShut
  {NoStop}%
\bibitem [{\citenamefont {Pekola}\ and\ \citenamefont
  {Karimi}(2021)}]{Pekola.21.RMP}%
  \BibitemOpen
  \bibfield  {author} {\bibinfo {author} {\bibfnamefont {J.}~\bibnamefont
  {Pekola}}\ and\ \bibinfo {author} {\bibfnamefont {B.}~\bibnamefont
  {Karimi}},\ }\bibfield  {title} {\enquote {\bibinfo {title} {Colloquium:
  Quantum heat transport in condensed matter systems},}\ }\href {\doibase
  10.1103/RevModPhys.93.041001} {\bibfield  {journal} {\bibinfo  {journal}
  {Rev. Mod. Phys.}\ }\textbf {\bibinfo {volume} {93}},\ \bibinfo {pages}
  {041001} (\bibinfo {year} {2021})}\BibitemShut {NoStop}%
\bibitem [{\citenamefont {Yuan}\ \emph {et~al.}(2022)\citenamefont {Yuan},
  \citenamefont {Ma},\ and\ \citenamefont {Sun}}]{Yuan.22.PRE}%
  \BibitemOpen
  \bibfield  {author} {\bibinfo {author} {\bibfnamefont {H.}~\bibnamefont
  {Yuan}}, \bibinfo {author} {\bibfnamefont {Y.-H.}\ \bibnamefont {Ma}}, \ and\
  \bibinfo {author} {\bibfnamefont {C.}~\bibnamefont {Sun}},\ }\bibfield
  {title} {\enquote {\bibinfo {title} {Optimizing thermodynamic cycles with two
  finite-sized reservoirs},}\ }\href {\doibase 10.1103/PhysRevE.105.L022101}
  {\bibfield  {journal} {\bibinfo  {journal} {Phys. Rev. E}\ }\textbf {\bibinfo
  {volume} {105}},\ \bibinfo {pages} {L022101} (\bibinfo {year}
  {2022})}\BibitemShut {NoStop}%
\bibitem [{\citenamefont {Spiecker}\ \emph {et~al.}(2023)\citenamefont
  {Spiecker}, \citenamefont {Paluch}, \citenamefont {Gosling}, \citenamefont
  {Drucker}, \citenamefont {Matityahu}, \citenamefont {Gusenkova},
  \citenamefont {G{\"u}nzler}, \citenamefont {Rieger}, \citenamefont
  {Takmakov}, \citenamefont {Valenti}, \citenamefont {Winkel}, \citenamefont
  {Gebauer}, \citenamefont {Sander}, \citenamefont {Catelani}, \citenamefont
  {Shnirman}, \citenamefont {Ustinov}, \citenamefont {Wernsdorfer},
  \citenamefont {Cohen},\ and\ \citenamefont {Pop}}]{Spiecker.23.NP}%
  \BibitemOpen
  \bibfield  {author} {\bibinfo {author} {\bibfnamefont {M.}~\bibnamefont
  {Spiecker}}, \bibinfo {author} {\bibfnamefont {P.}~\bibnamefont {Paluch}},
  \bibinfo {author} {\bibfnamefont {N.}~\bibnamefont {Gosling}}, \bibinfo
  {author} {\bibfnamefont {N.}~\bibnamefont {Drucker}}, \bibinfo {author}
  {\bibfnamefont {S.}~\bibnamefont {Matityahu}}, \bibinfo {author}
  {\bibfnamefont {D.}~\bibnamefont {Gusenkova}}, \bibinfo {author}
  {\bibfnamefont {S.}~\bibnamefont {G{\"u}nzler}}, \bibinfo {author}
  {\bibfnamefont {D.}~\bibnamefont {Rieger}}, \bibinfo {author} {\bibfnamefont
  {I.}~\bibnamefont {Takmakov}}, \bibinfo {author} {\bibfnamefont
  {F.}~\bibnamefont {Valenti}}, \bibinfo {author} {\bibfnamefont
  {P.}~\bibnamefont {Winkel}}, \bibinfo {author} {\bibfnamefont
  {R.}~\bibnamefont {Gebauer}}, \bibinfo {author} {\bibfnamefont
  {O.}~\bibnamefont {Sander}}, \bibinfo {author} {\bibfnamefont
  {G.}~\bibnamefont {Catelani}}, \bibinfo {author} {\bibfnamefont
  {A.}~\bibnamefont {Shnirman}}, \bibinfo {author} {\bibfnamefont
  {A.}~\bibnamefont {Ustinov}}, \bibinfo {author} {\bibfnamefont
  {W.}~\bibnamefont {Wernsdorfer}}, \bibinfo {author} {\bibfnamefont
  {Y.}~\bibnamefont {Cohen}}, \ and\ \bibinfo {author} {\bibfnamefont
  {I.}~\bibnamefont {Pop}},\ }\bibfield  {title} {\enquote {\bibinfo {title}
  {Two-level system hyperpolarization using a quantum szilard engine},}\ }\href
  {\doibase 10.1038/s41567-023-02082-8} {\bibfield  {journal} {\bibinfo
  {journal} {Nat. Phys.}\ }\textbf {\bibinfo {volume} {19}},\ \bibinfo {pages}
  {1320} (\bibinfo {year} {2023})}\BibitemShut {NoStop}%
\bibitem [{\citenamefont {Moreira}\ \emph {et~al.}(2023)\citenamefont
  {Moreira}, \citenamefont {Samuelsson},\ and\ \citenamefont
  {Potts}}]{Moreira.23.PRL}%
  \BibitemOpen
  \bibfield  {author} {\bibinfo {author} {\bibfnamefont {S.}~\bibnamefont
  {Moreira}}, \bibinfo {author} {\bibfnamefont {P.}~\bibnamefont {Samuelsson}},
  \ and\ \bibinfo {author} {\bibfnamefont {P.}~\bibnamefont {Potts}},\
  }\bibfield  {title} {\enquote {\bibinfo {title} {Stochastic thermodynamics of
  a quantum dot coupled to a finite-size reservoir},}\ }\href {\doibase
  10.1103/PhysRevLett.131.220405} {\bibfield  {journal} {\bibinfo  {journal}
  {Phys. Rev. Lett.}\ }\textbf {\bibinfo {volume} {131}},\ \bibinfo {pages}
  {220405} (\bibinfo {year} {2023})}\BibitemShut {NoStop}%
  \bibitem [{\citenamefont {Strasberg}\ \emph {et~al.}(2021)\citenamefont
  {Strasberg}, \citenamefont {D\'{\i}az},\ and\ \citenamefont
  {Riera-Campeny}}]{Strasberg.21.PRE}%
  \BibitemOpen
  \bibfield  {author} {\bibinfo {author} {\bibfnamefont {P.}~\bibnamefont
  {Strasberg}}, \bibinfo {author} {\bibfnamefont {M.}~\bibnamefont
  {D\'{\i}az}}, \ and\ \bibinfo {author} {\bibfnamefont {A.}~\bibnamefont
  {Riera-Campeny}},\ }\bibfield  {title} {\enquote {\bibinfo {title} {Clausius
  inequality for finite baths reveals universal efficiency improvements},}\
  }\href {\doibase 10.1103/PhysRevE.104.L022103} {\bibfield  {journal}
  {\bibinfo  {journal} {Phys. Rev. E}\ }\textbf {\bibinfo {volume} {104}},\
  \bibinfo {pages} {L022103} (\bibinfo {year} {2021})}\BibitemShut {NoStop}%
\end{thebibliography}
%

\begin{widetext}

{\Large{\bf Supplemental material:} Supplemental Material: Global-Local Duality of Energetic Control Cost in Multipartite Quantum Correlated Systems}
\\
\\
\\

In this supplemental material, we first analyze two special scenarios: (i) vanishing interaction between local parties, and (ii) the reversible limit. We then present additional simulation results that complement those showed in the main text. We also discuss the possibility of generalizing the main results showed in the main text to account for finite-sized reservoirs.

\section{I.~Vanishing interaction between local parties}\label{sec:2}
In this section we demonstrate that the global and local costs equal when there is no physical interaction, namely, $H_{\rm{I}}=0$. With the work definition $W_{g}(t)\equiv\int_0^t\mathrm{Tr}[\rho_{g}(\tau)\frac{d}{d\tau}H_{g}(\tau)]d\tau$ for the global system, we readily infer $W_{g}(t)= \sum_iW_i(t)$ with $W_{i}(t)\equiv\int_0^t\mathrm{Tr}[\rho_{i}(\tau)\frac{d}{d\tau}H_{i}(\tau)]d\tau$ from the Hamiltonian form $H_g(t)=\sum_iH_i(t)$. Hence in this particular scenario the dissipated work contrast $W_{\rm{dis}}^{\Delta}(t)$ defined in the main text reduces to 
\bea
   W_{\rm{dis}}^{\Delta}(t) &=& \sum_{i}\Delta F_i(t)-\Delta F_{g}(t)\nonumber\\
        &=& \sum_{i}\Delta E_{i}(t)-\Delta E_{g}(t)\nonumber\\
        &&-T[\sum_{i}\Delta S_{i}(t)-\Delta S_{g}(t)]\nonumber\\
        &=& -T\Delta I_g(t).
\eea
To arrive at the last line, we have noted the fact that $\Delta E_{g}(t)=\sum_{i}\Delta E_{i}(t)$ with $E_g(t)=\mathrm{Tr}[H_g(t)\rho_g(t)]$ and $E_i(t)=\mathrm{Tr}[H_i(t)\rho_i(t)]$ in the absence of physical interaction. We have also utilized the definition of quantum multipartite mutual information $I_g(t) \equiv \sum_i S_i(t)-S_{g}(t)$ \cite{Modi.12.RMP,Huber.15.NJP,Chiara.18.RPP}. Inserting the above expression into the sum rule showed in the main text, we get 
\bea
    Q_g(t) &=& Q_l(t) +T\Delta I_g(t)-T\Delta I_g(t)\nonumber\\
        &=& Q_l(t).
\eea
Hence, we show that the global and local cost are just the same when there is no coupling between local parties. Besides this particular case, we generally expect that $Q_g(t)\neq Q_l(t)$. 

\section{II.~The reversible limit}\label{sec:3}
For the global system weakly coupled to a thermal bath at a temperature $T$, it should obey the standard form of the second law of thermodynamics
\begin{equation}\label{eq:sl}
    T\Delta S_{g}(t)+Q_g(t)~\geqslant~0.
\end{equation}
Combining the above Clausius inequality with the energy decomposition $\Delta E_{g}(g)= -Q_g(t)+W_{g}(t)$ and noting the change in the global nonequilibrium free energy $\Delta F_{g}(t) = \Delta E_{g}(t)-T\Delta S_{g}(t)$, we arrive at the principle of maximum work for the global system
\begin{equation}\label{eq:mw}
    W_g(t)-\Delta F_g(t)~\geqslant~0.
\end{equation}
For later convenience, we also refer to Eq. (\ref{eq:mw}) as a standard form of the second law of thermodynamics. The equality is attained in the reversible limit, implying that the dissipated work of the global system $W_{\rm{dis}}^g(t)=W_g(t)-\Delta F_g(t)$ vanishes in that limit.

Now we try to show that subsystems of a global correlated one do not satisfy the standard forms of the second law of thermodynamics aforementioned such that the local dissipated work $W_{\rm{dis}}^l(t)=\sum_i[W_i(t)-\Delta F_i(t)]$ remains nonzero in the reversible limit. For simplicity, we focus on a bipartite correlated system to carry out the derivation. Extension to a multipartite correlated system is straightforward. Combining the energy decomposition of the global system $\Delta E_{g}(t)=\sum_{i=1}^2\Delta E_{i}(t)+\Delta E_{\rm I}(t)$ ($E_{\rm I}(t)=\mathrm{Tr}[H_{\rm I}(t)\rho_g(t)]$) according to the form of the global Hamiltonian with changes in local nonequilibrium free energies $\Delta F_{i}(t) = \Delta E_{i}(t)-T\Delta S_{i}(t)$ ($i=1, 2$), we receive the following relation
\bea
   \Delta E_{g}(t) &=& \sum_{i=1}^2[\Delta F_{i}(t)+T\Delta S_{i}(t)] +\Delta E_{\rm I}(t)\nonumber\\
        &=&-Q_g(t) +W_{g}(t).
\eea
From the above equation, we can get an expression for the global cost $Q_g(t)=W_g(t)-\sum_{i=1}^2[\Delta F_{i}(t)+T\Delta S_{i}(t)]-\Delta E_{\rm I}(t)$. Inserting this expression into the left-hand-side of Eq. (\ref{eq:sl}) and noting the definition of the quantum mutual information $I(t)$, we get the following inequality  
\begin{equation}\label{eq:local_sl}
   \sum_{i=1}^2[ W_{i}(t)-\Delta F_i(t)]~\geqslant~ T\Delta I (t)+\Delta E_{\rm I}(t).
\end{equation}
In arriving at the above inequality, we have noted the fact that $W_g(t)=\sum_{i=1}^2W_{i}(t)$ as only the subsystem parts are being driven. A direct indication of the above inequality is that the subsystems do not satisfy the standard form of the second law of thermodynamics which instead states that $W_{i}(t)-\Delta F_{i}\geqslant 0$ ($i=1,2$). By further noting that the left-hand-side of Eq. (\ref{eq:local_sl}) is just the local dissipated work $W_{\rm{dis}}^l(t)$ we define in the main text, we thus have in the reversible limit that
\begin{equation}
    W_{\rm{dis}}^l(t)~=~T\Delta I (t)+\Delta E_{\rm I}(t)
\end{equation}
which remains nonzero in general. Consequently, $W_{\rm{dis}}^{\Delta}(t)$ remains finite in the reversible limit.

\section{III.~Additional numerical results}\label{sec:4}
\subsection{A.~Numerical results for $F^{\Delta}(t)$ and $W^{\Delta}(t)$}
\begin{figure}[tbh!]
 \centering
 \includegraphics[width=0.65\columnwidth]{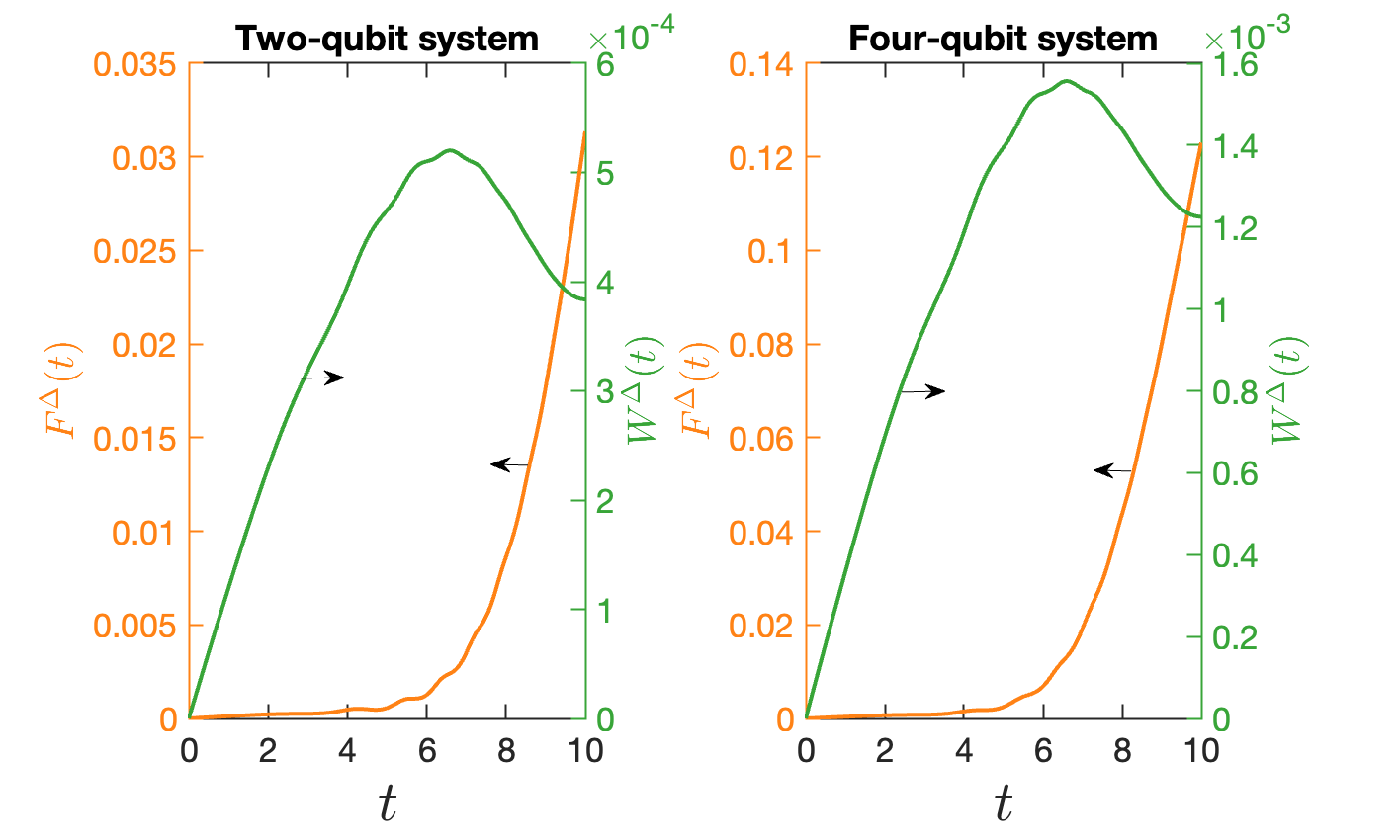} 
 \caption{Time-dependence of $F^{\Delta}(t)$ (left axis) and $W^{\Delta}(t)$ (right axis) for models studied in the main text. Left panel: Two-qubit system. Right panel: Four-qubit system. Parameters are $\beta=1$, $\lambda=0.02$, $\gamma=0.02$, $\varepsilon_0=0.4$, $\varepsilon_{\tau}=10$, $N=4$ and $\tau=10$.}
\protect\label{fig:data2}
\end{figure}
In Fig. \ref{fig:data2}, we depict numerical results for $F^{\Delta}(t)=\Delta F_g(t)-\sum_i\Delta F_i(t)$ and $W^{\Delta}(t)=W_g(t)-\sum_iW_i(t)$ that complement those shown in the main text. As can be seen from both figures, $F^{\Delta}(t)$ and $W^{\Delta}(t)$ are positive under the selected control protocols, and particularly, $W^{\Delta}(t)$ is at least one order of magnitude smaller than $F^{\Delta}(t)$.

\subsection{B.~Multi-qubit model with $Q_g(t)<Q_l(t)$}
\begin{figure}[tbh!]
 \centering
 \includegraphics[width=0.65\columnwidth]{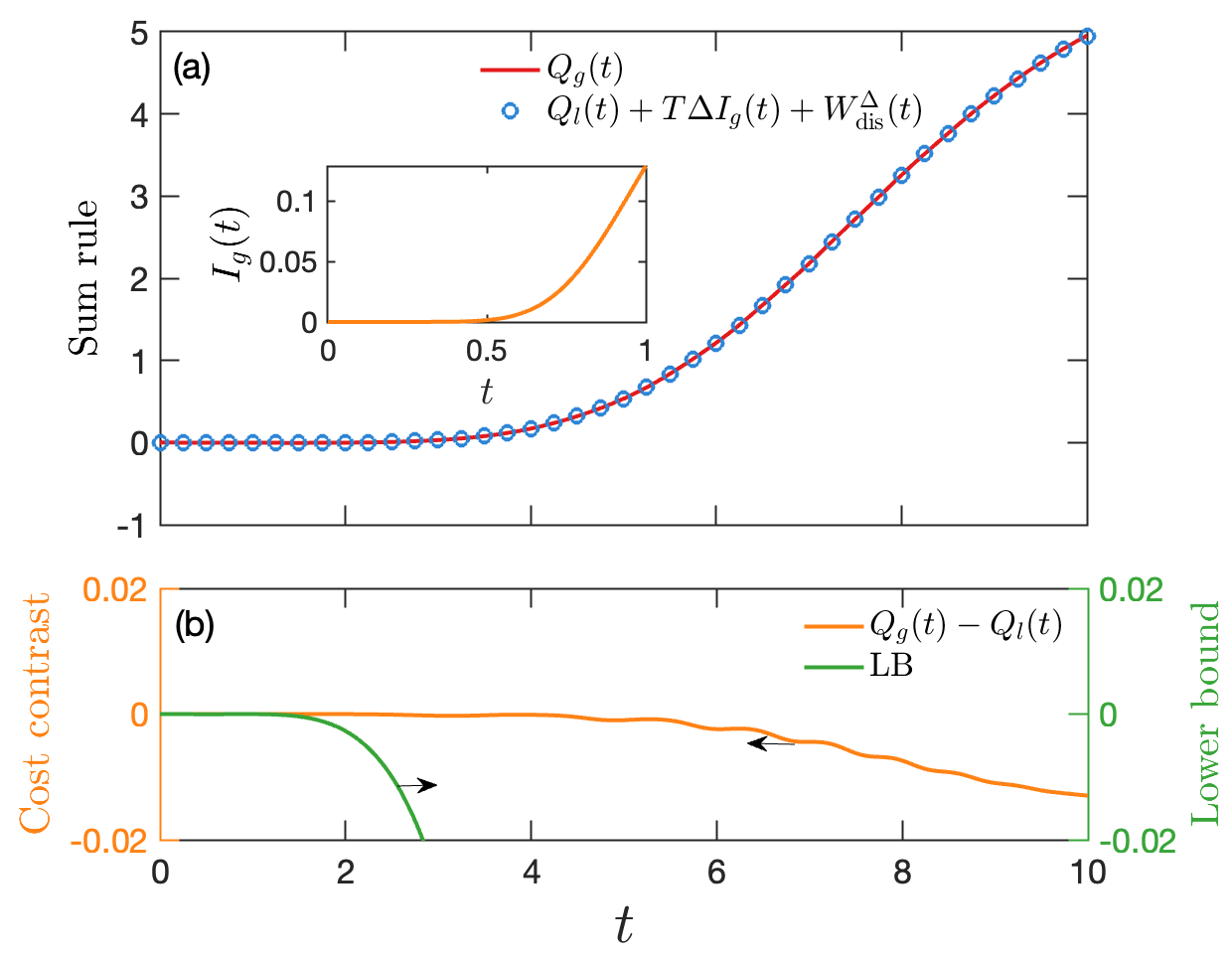} 
 \caption{Behaviors of thermodynamic cost of erasing an open multiqubit system described by Eq. (\ref{eq:HS_e}). (a) Validity of the sum rule for the global cost at finite times. Inset: Time-dependence of quantum multipartite mutual information $I_g(t)$. (b) The cost contrast $Q_g(t)-Q_l(t)$ (solid line, left axis) and its lower bound (LB) [Eq. (3) in the main text] (dashed line, right axis). Parameters are $\beta=1$, $\lambda=0.02$, $\gamma=0.02$, $\varepsilon_0=0.4$, $\varepsilon_{\tau}=10$, $N=4$ and $\tau=10$.}
\protect\label{fig:data1}
\end{figure}
As elaborated in the main text, the cost contrast $Q_g(t)-Q_l(t)$ does not have a definite sign. We have depicted numerical examples showing $Q_g(t)-Q_l(t)>0$ by using multi-qubit models in the main text. Here, we further show that a result of $Q_g(t)-Q_l(t)<0$ can also occur in multi-qubit systems.

Specifically, we consider the following multi-qubit system with a different intra-qubit interaction form, $\lambda_t \sigma_{i}^{z}\sigma_{i+1}^{z}$, compared to that in the main text. The Hamiltonian of the global system is now represented as 
\begin{equation}\label{eq:HS_e}
    H_g^{'}(t) = \frac{\varepsilon_t}{2}\sum_{i=1}^N\left[\cos (\theta_t)\sigma_i^{z}+\sin(\theta_t)\sigma_i^{x}\right]+\sum_{i=1}^{N-1}\lambda_t \sigma_{i}^{z}\sigma_{i+1}^{z}.
\end{equation}
Besides the change in the intra-qubit interaction form, other terms and notations in the above Hamiltonian remain the same with those considered in the main text, including the control protocols. The quantum Lindblad master equation governing the evolution of $\rho_g(t)$ still adopts the standard form illustrated in the main text with just the Hamiltonian part being replaced by Eq. (\ref{eq:HS_e}).

A typical set of numerical results obtained using an initial thermal state for the global system $\rho_{g}^{'}(0)= e^{-\beta H_g^{'}(0)}/tr[e^{\beta H_g^{'}(0)}]$ is depicted in Fig. \ref{fig:data1}. Comparing with numerical results shown in the main text, we reaffirm the validity of the sum rule in Fig. \ref{fig:data1} (a), which consistently describes the relationship between the thermodynamic cost and the total correlation as well as dissipated work during the finite-time evolution process in a correlated system. However, a notable difference is observed in Fig. \ref{fig:data1} (b), where $Q_g(t)$ becomes instead smaller than $Q_l(t)$ at large times. This behavior robustly substantiates our previous assertion that the cost contrast $Q_g(t) - Q_l(t)$ does not have a definite sign; different model configurations can yield distinct outcomes. Nonetheless, it is evident that the lower bound still constrains the cost contrast from the below, albeit somewhat loosely.

\section{IV.~Accounting for finite-sized reservoirs}\label{sec:5}
To account for finite-sized reservoirs described by a time-dependent effective temperature $T(t)$ \cite{Schaller.14.NJP,Amato.20.PRA,Yuan.22.PRE,Campeny.21.PRXQ}, we consider a generalized form of the nonequilibrium free energy \cite{Liu.23.PRAa}. For the global system, we now have
\begin{equation}
   \mathcal{F}_{g}(t)~=~E_{g}(t)-T(t)S_{g}(t). 
\end{equation}
Combining the first law of thermodynamics for the global system $\Delta E_g(t)=W_g(t)-Q_g(t)$ and the definition of quantum multipartite mutual information $I_g(t)$ \cite{Modi.12.RMP,Huber.15.NJP,Chiara.18.RPP}, the change in $\mathcal{F}_g(t)$ can be expressed as 
\begin{equation}
    \Delta \mathcal{F}_{g}(t)=W_{g}(t)-Q_g(t)-T(t)[\sum_i S_i(t)-I_g(t)]+T(0)[\sum_i S_i(0)-I_g(0)].
\end{equation}
To proceed, we further consider generalized nonequilibrium free energies $\mathcal{F}_i(t)~=~E_i(t)-T(t)S_i(t)$ for local parties. We can use $\mathcal{F}_i(t)$ twice at both time $0$ and $t$ to replace the terms $\sum_i S_i(t)$ and $\sum_i S_i(0)$ in $\Delta \mathcal{F}_g(t)$, yielding
\bea\label{eq:ddf_e}
    Q_g(t) &=& W_{g}(t)-\Delta \mathcal{F}_{g}(t)+T(t)I_g(t)-T(0)I_g(0)\nonumber\\
    &&+\sum_i[\Delta \mathcal{F}_i(t)-\Delta E_i(t)].
\eea
Invoking the energy decomposition for local parties $\Delta E_i(t)=W_i(t)-Q_i(t)$ and introducing generalized definitions of dissipated work
\bea
   \mathcal{W}_{\rm{dis}}^{g}(t) &\equiv& W_g(t)-\Delta\mathcal{F}_g(t),\nonumber\\
   \mathcal{W}_{\rm{dis}}^{l}(t) &\equiv& \sum_i[W_i(t)-\Delta\mathcal{F}_i(t)],
\eea
we can obtain from Eq. (\ref{eq:ddf_e}) the following generalized thermodynamic cost sum rule
\begin{equation}\label{eq:sl_e}
    Q_g(t)~=~Q_l(t)+T(t)I_g(t)-T(0)I_g(0)+\mathcal{W}_{\rm{dis}}^{\Delta}(t)
\end{equation}
with $\mathcal{W}_{\rm{dis}}^{\Delta}(t)=\mathcal{W}_{\rm{dis}}^{g}(t)-\mathcal{W}_{\rm{dis}}^{l}(t)$. Comparing Eq. (\ref{eq:sl_e}) with the sum rule showed in the main text, we know that the basic form remains even in the presence of finite-sized reservoirs with the change in the quantum multipartite mutual information now accounting for the time-dependence of an effective temperature. 

As for the cost contrast $Q_g(t)-Q_l(t)$, we remark that a simple extended form of the lower bound generally does not exist. Following the proof details showed in the main text that leads to a general lower bound, we at most obtain
\begin{equation}\label{s14}
    T(t)S_g(t)-T(0)S_g(0)~=~-Q_l(t)+\mathcal{W}_{\rm{dis}}^g(t)+\Delta E_{\rm I}(t).
\end{equation}
In the presence of time-dependent temperature, the Clausius inequality takes the following intricate form \cite{Strasberg.21.PRE}
\begin{equation}\label{eq:ci}
    \Delta S_g(t)+\int\frac{dQ_g(t)}{T(t)}~\ge~0.
\end{equation}
Since the integral in Eq. (\ref{eq:ci}) cannot be expressed as a simple difference in heat dissipation between time $t$ and $0$, we generally cannot combine the above two equations Eqs. (\ref{s14}) and (\ref{eq:ci}) to get a lower bound on the cost contrast $Q_g(t)-Q_l(t)$ that resembles the one showed in the main text.

\end{widetext}

\end{document}